# Understanding Error Correction and its Role as Part of the Communication Channel in Environments composed of Self-Integrating Systems

*Aleksander Lodwich*

*aleksander[at]lodwich.net*

**Abstract – The raise of complexity of technical systems also raises knowledge required to set them up and to maintain them. The cost to evolve such systems can be prohibitive. In the field of Autonomic Computing, technical systems should therefore have various self-healing capabilities allowing system owners to provide only partial, potentially inconsistent updates of the system. The self-healing or self-integrating system shall find out the remaining changes to communications and functionalities in order to accommodate change and yet still restore function. This issue becomes even more interesting in context of Internet of Things and Industrial Internet where previously unexpected device combinations can be assembled in order to provide a surprising new function. In order to pursue higher levels of self-integration capabilities I propose to think of self-integration as sophisticated error correcting communications. Therefore, this paper discusses an extended scope of error correction with the purpose to emphasize error correction's role as an integrated element of bi-directional communication channels in self-integrating, autonomic communication scenarios.**

## 1 Motivation

Today, Internet of Things (IoT) is understood as ubiquitous presence of Internet-enabled devices [1]. The promise of IoT is that devices can massively interact among each other by relying on ubiquitous presence of a global communication channel: The Internet.

In practice, however, the vision suffers several problems, some more and some less obvious.

The first problem is that communication between devices must be technically compatible. Individual service and protocol implementations using the Internet usually lack the necessary level of compatibility and interop-erability in order to convince a broad user base. Interoperability platforms like Thing-Worx[1] were started in order to overcome a wide range of typical communication challenges but are not widely accepted. Whether such broad acceptance can occur, will depend on Killer-applications demonstrated in industry and consumer sphere. Currently, best business opportunities offered by IoT technology are maintenance and fast adaptation of industrial facilities where the Ethernet is replacing many other networking technologies[2] and allows continuous expert maintenance and surveillance by the OEM. Improved connectivity shall not only improve interoperability of industrial facilities but also make them more elastic, auto-configurating and resilient to errors [2]. Industrial cost advantages attained in this way are currently and in near future not reproducible for consumers. It is unclear how consumers can benefit from IoT and technologies developed mainly for the industry..

The second problem is data security. Currently marketed solutions usually try to deposit data in a "cloud" where users can exploit and monitor complex technological environments. Companies running these clouds collect critical data about the customer and can hopefully offer more comfortable features in exchange. However, after Snowden reveled the scope of surveillance practiced by states, concerns about personal freedom and about keeping options for effective civil resistance against state power has been even raised. It depends on any individual's calculation of the matter whether technical conveniences outweigh diminished political sovereignty. I speculate, the more abuse is practiced by states the lesser will be the acceptance of any cloud-based technology.

The third problem is that companies enable their products for the Internet in order to sell their *own* products. It is currently not clear how the so enabled products will allow business-on-business practice for third parties. For example, such business model could al-low to buy a 3rd-party plug-in for an IoT-en-abled multi-functional printer (with scanner) in order to enhance it with OCR capabilities. Although business-on-business is propagated along with IoT, only few concrete examples exist in order to demonstrate such capabilities. The classic smart phone is currently the only real Internet-enabled platform which is clearly supporting 3rd-party business models.

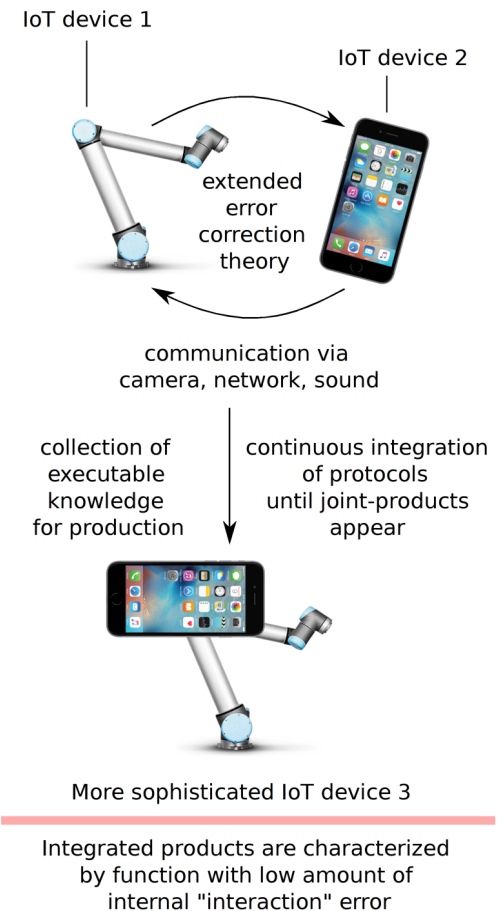

*Figure 1: Two IoT-devices could be assembled and upgraded in order to perform an unexpected new function.*

The fourth problem, which is often overlooked, is that Internet of Things is not about the Internet. The main idea behind IoT is about flexible object interaction [3] as suggested in figure 1. For such interaction any number of communication channels are imaginable and in fact, the Internet could play the least role in it. Products are constantly extended with sensors and additional actors for convenience functions and products of the fu-ture might be shipped with functionality which cannot be understood prior to understanding democratized production concepts. By delivering consumer products which are more general-purpose (like the smart-phone) and which are equipped with hardware and communication technology (such as for example Bluetooth) not necessary for bare function, customers receive more opportunities to automate their living environment and to provide non-commercial conveniences. This may require hardly any Internet connectivity, if any at all. Nevertheless, home Ethernet networks can play an important supportive role in realizing such functions, for example stretching communications from room to room or from household to household (without any intermediating parties). Even better, functionality developed in such an environment could become a portable item in sense of a product.

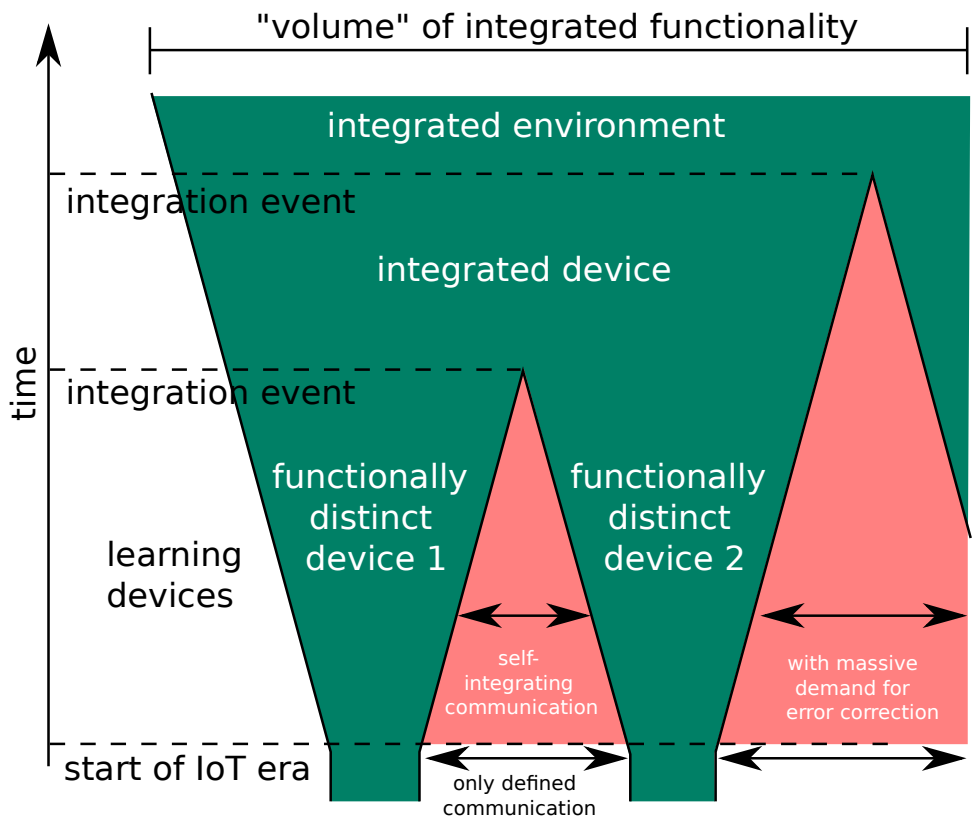

*Figure 2:Role of communications in self-integrating environments. At first, functionally distinct devices try to communicate with each other in order to perform an unknown task. Those devices were not designed for the task and share only limited amount of shared knowledge. By enabling sophisticated error correction mechanisms the devices learn to identify and translate between their internal concepts and received signals. This process is concerned with elimination of systematic errors and hence cannot be understood in sense of error correction designed to counteract random error sources.*

In any case, visual, acoustic or other human-machine-machine interaction channels become more dominant in such scenarios.

In the following rest of the paper I will concentrate on the idea that IoT is currently failing to convince on the fourth challenge: Ability to integrate devices for customer-sought purposes where a hybrid constellation of communication technologies is involved. In such scenarios systems must at first pick up loose communications and constantly enhance them in order to gain communication precision while at the same time growing the number of functions which they can communicate about (cf. figure 2).

This process starts with communications containing large amounts of redundancy or overhead which is used by the communication parties in order to retrieve parts of the communicated information. As the communica-tion parties evolve, communications become more concise and require less corrective information to be provided. Ultimately, systems adapt to each other to such a degree that systematic errors in communications are eliminated. These points of convergence are marked in figure 2 as *integration events*. As of this point the system is functionally integrated and error correction only remains in place to repair random errors.

In self-integrating environments systematic errors can appear as random errors and hence they require integrated approaches. However, most error correction literature [4][5][6] motivates error correction with random sources of error as is shown in figure 3. Modern error correction algorithms are all designed to provide additional noise robustness by transmitting redundant information which is used by the receiver to reverse-conclude original signal.

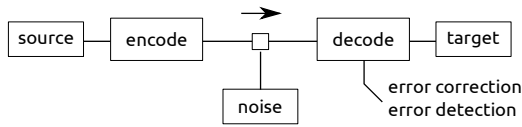

*Figure 3: Standard model of communication and forward error detection / correction.*

In this paper I will explore a more general understanding of error correction which includes systematic errors, error correction in nested channels and dynamic error correction relying on dynamical optimization of representations (codes, modulations). This expanded scope is touching upon concepts of Autonomic Computing [7], [8] and some advanced capabilities associated with it such as *self-healing*, *self-stabilization* or *self-calibration* [9]. With the extended view I would like to investigate how error correction in communications would translate into structural error correction and vice versa, i.e. if structural failures can be translated into communication errors and if communication correction can restore structures. Structural argumentation is important because technical functionality is organized in architectures which must be maintained.

I will also avoid an all too quick concentration on digital communications when speaking of error correction can obfuscate opportu-nities to use error correction techniques in new domains or in innovative ways. Let us think of two examples where one would not think of error correction right away:

Example 1: Input helpers on mobile phones are an example of non-binary (correction of words) error correction technique allowing the user the increase input bandwidth at cost of input reliability. In fact, keyboards like the Swype keyboard allow to relax requirements on input protocols even further (such as producing distinct input events by pressing character buttons). Only sophisticated error correcting communication channels can tolerate the amount of errors produced during the fast interaction between users and mobile phones.

Example 2: Binarization thresholds used to detect digital signal levels could depend on their position in a protocol or expected higher level semantics. This would suppress sampling of unlikely values which are associated with errors (speed is an issue here, though).

The above examples are examples from different domains of technology and could be considered either as issues of user friendliness or signal processing. However, I believe that such technologies are better understood as error correction features in communications.

An integrated concept of error correction holds the promise to create robust communications which can involve machines and humans as endpoints alike. This shall be achieved by better allocating the question of reliability and systematic deviations to different protocol levels. Current communication technology requires a comprehensive compatibility of protocol stacks for correct function. This can exceed the possibilities of outdated systems, unrelated systems or involved humans.

This paper will emphasize the nested nature of error correction for any type of communication, be this using digital, analog, virtual, physical, scalar or complex signals, and to suggest that more intelligent error correction in terms of self-learning and self-elimination could be possible than is provided today. In fact, I believe that by gaining better understanding of error correction as a systemic property will also improve design of many

APIs and gadget interplay.

This gadget interplay is heavily relying on devices to implement extremely strong error correcting channels in order to be usable in unprecedented ways (*user driven innovation*). Achieving this will require a more complex view of error correction which establishes relationship between error correction, pattern recognition, knowledge-based reasoning, protocol stack design and control theory.

## 2 Related Research

### 2.1 Scope of Related Research

This work follows the question how systems could discover communication partners and iteratively improve communication channel performance with them in the sense that any kind of communication error is eliminated dynamically. It is my hypothesis that such dynamic error correction will result in self-integration of systems because integrated systems are characterized by absence of systematic communication errors and by rich interactions.

However, dynamic error correction is only one aspect of self-integration as integration also integration of function concepts. Nevertheless, understanding main concepts behind dynamic error correction is absolutely necessary in order to explain how self-configuring systems denominated in omega-units [10] could strive for higher degrees of integration.

In the following subsections I will thus highlight several research directions which are related to dynamic error correction in systems.

### 2.2 Classic Digital Codes and Analog Modulations

Error correcting codes are the first domain of research which comes to mind when thinking of error correction in communication channels. Literature concerned with error correcting codes (such as [4]–[6]) is not covering the dynamical process of spawning communications and developing robust protocols of communication including the necessary robust codes. Hamming codes, low-density codes, block codes etc. assume random noise sources on digital channels and a correction method to deal with it right from the beginning.

The different methods differ in the methodology how they generate and distribute redundancy during encoding and how this information is used to recover original messages (forward error correction). However, a common ground to understand those methods is to see them as mechanisms to generate directed graphs between legal codes and illegal codes, and where each transition points towards the most likely legal code (as shown for example in figure 4).

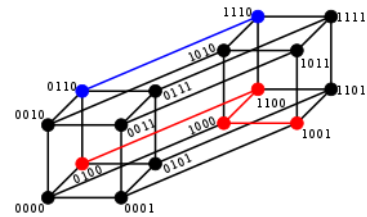

*Figure 4: A graph in shape of a 4D cube: Three arcs connect between any legal code. Moving away one arc away by error can be corrected.[3]*

Another equivalent way to see their operation is to distribute legal codes in a higher than necessary space and then to classify any received value to the nearest legal code. Because the spaces of operation are usually binary (best suitable for computers), methods used for error detection and correction in digital communications are made particularly efficient in such spaces, sometimes at cost of generality. Most algorithms are not easily convertible to non-digital codes.

In analog communications error correction is understood as robust modulation of signal [11]. Such signal can travel long distances and accumulate a lot of distortion until the receiver is unable to interpret it correctly. In this paper I will treat modulation and error robust digital encoding in a unified manner in sense of space-time codes as is presented in [12].

In digital communications error correction is understood as choosing a single codeword from a very large code set. This code set can become intractable in which case correction capabilities are restricted down to the point where errors can be only detected. Received digital messages can be seen a data structures which can be monitored for corruption in memory [13].

---

3 https://en.wikipedia.org/wiki/Hamming_distance

There exist approaches trying to mix the two worlds: For example, Turbo Codes are a mix of digital coding and analog modulation techniques ("soft coders") explicitly trying to combine advantages of continuous values in binary communications [14]. Such hybrid methods best demonstrate that it is useful to view error correction and detection technology from a broader perspective.

### 2.3 Artificial Agents and Robotics

Mobile robotics are a very similar field of application as IoT devices and we will probably see a merging of the two in the future. Robots must be able to solve the task at hand either all by themselves or they must collaborate in order to achieve the mission goals. Of particular interest in context of self-integrating systems is research that is demonstrating how robots can detect other robots in their environment and learn to communicate and to organize with them.

In this context, Cristiano Castelfranchi describes in [15] the concept of *behavioral implicit communication* – a concept for understanding various ways that systems can interact in more or less intentional ways and by conveying information they can cooperate or harm each other. In technical systems we usually take benevolent communications as given (i.e. systems are "sincere and helpful" or "altruistic" [16], [17]) but in scenarios where systems want to integrate with yet unknown devices, motivations and risk of partners' strategies must be considered.

Since agents can interact with each other in different modes (help, control, destroy, block, subordinate, etc.) the nature of communications and chosen channel characteristics will depend on the type of interaction. In figure 5 I have simplified the different interactions to just four sectors, similarly to the model reported in [18] .

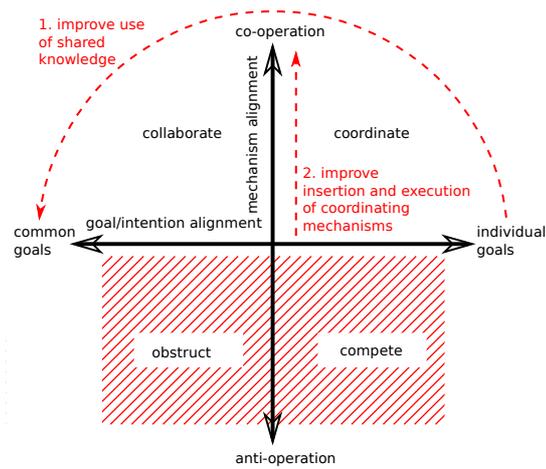

*Figure 5: Systems can collaborate, coordinate, obstruct or compete with each other. (Adapted version from [18])*

There are two dimensions in the above figure: horizontal dimension is used to describe alignment between intention and the vertical dimension describes alignment of mechanisms. Systems with aligned goals and intentions can either collaborate or obstruct each other depending whether their mechanisms were aligned or not. If systems have individual (contrary) intentions they can still either coordinate or compete. Competition is the destructive part where systems engage in communications and interactions which shall be harmful to the recipient. However, systems with competing intentions still can align their mechanisms in order to coordinate their activities. For example if two cars want to go through a narrow bridge then they can either coordinate their travel with a semaphore (obedience to semaphore is the mechanism alignment) or try to push the other car backwards (compete for way).

Dynamic error correction's job would be to align those mechanisms until integration is achieved [19]. According to this model, integration will demand that systems' relationships have been moved out of the dashed area.

According to Castelfranchi, systems must learn to interpret and understand other systems' normal behavior and to detect special behaviors in order detect meaningful signals usable for communication. However, this approach is rare in research. Most designers choose to provide dedicated, "explicit" communication channels for connecting and coordinating robot hives (examples are [20]–[23]).

*"Existing coordination methods are mainly based on the use of explicit communication." [23]*

There are examples of research which is investigating the process of learning communication. Stefano Nolfi has shown and discussed in [24] more concretely how embodied agents (robots) can pickup simple communications by observing body signals and acoustic signals. Nolfi and others [18], [25]–[28] have found it difficult to clearly separate communication from non-communicative actions or Stigmergy [29]. According to Nolfi this is accompanied by a controversy how to define communication.

Therefore the problem of continuous development of communication is treated in chapter 10 (pp. 34) in such a way as to show that there is a continuum between functional and communicative behaviors and that stigmergical communications lie in between two endpoints of a spectrum. I will rely on the idea that a communication channel is an entanglement of configuration spaces of two or more objects and that systems can only learn to use them if they have internal states useful for being communicated, there is self-monitoring, environment modeling and a sophisticated auto-resonant, error-correcting behavior. I believe that this approach is capable to reconcile observed data, technical solutions and critique found in literature because defining the terms is then reduced to just choosing convenient boundaries in such continuum. Of course, depending on the implemented capabilities of communicating systems, developed channel solutions will populate that spectrum more or less evenly.

By 2003 Luc Steels wrote that developing communications and interactions beyond pre-programmed ways is particularly difficult if systems must evolve more complex temporal structure of communication (grammar) [26]. Since grammar is a temporal pattern and often used with communication of graphical data I will superficially delve into developing spatio-temporal codes in communications and their relationship to graphical structures and protocols in chapters 4.3 and 5.

## 2.4 Error Correction in high-level Languages

In recent years, technology started to offer natural language input more ubiquitously. Human machine interfaces (HMI) based on natural language are particularly interesting for developing and analyzing advanced communication channel properties and error correction capabilities as the data is full of errors to deal with. Correction of input from interaction channels between humans users and devices is key to efficient and swift device operation. Input propositions made in search masks are just one broadly known example.

Communication via text streams is particularly unreliable on small devices with physically unsound user interfaces such as on mobile phones. Efficient input can demand simultaneous use of several tolerant techniques in order to be satisfactory for users [30] such as swiping [31], modal interfacing (switchable input method), contextual word propositions [32], spell and grammar checking [33], gestures or parallel support by voice. Deduction of the most likely input sequence is following a remarkably similar objective to what error correction is trying to achieve on binary data but can be far more multi-media.

Error correction for high-level languages can demand a large amount of a priori knowledge which is usually too large to be provided as additional information by the sender and hence is provided directly to the receiver's knowledge pool. Such a solution is described in [34]: The basis of the correction was a huge database of N-grams provided by Microsoft. The solution has chosen among similarly sounding words which fit better into the context of the sentence. Similar approaches have also been demonstrated for OCR [35] and speech [36].

In *natural language training* the term error correction refers to the attempt to improve a learner's communication competence by giving him feedback on its performance. Clearly, this term is coined more to the idea of a systematic error that a human speaker is making. Nevertheless, his errors materialize as proto-col ("grammar errors") and codeword-errors ("spelling errors" or "word choice errors") on the channel (text or speech).

Rozkovskaya and Roth applied the term error correction to optimize training in regard to context-sensitive spelling errors [37]. This work is formulated in context of language training but since this paper is interested in how communications generally emerge and stabilize between systems it seems to be worthwhile to take note of the idea that externally provided error correction is an established element in improving learners' language competence – an approach applicable to interaction between machines which would like to acquire a common protocol among them.

## 2.5 Other Areas of Error Correction

- DNA repair [38] can be considered as error correction for instruction blocks.
- The W3C consortium is pushing to establish omnipresence of OWL and RDF technologies in order to allow services to complete, correct and transform data based on suppliable pieces of knowledge about the communicated objects [39], [40].

## 2.6 What can be concluded from Current State of Research

Current state of research demonstrates that error correction is a prominent feature in many applications and in many cases a key success factor to product's popularity but named research does not relate to each other in terms of references to a common theory or concept. Instead engineers and researchers dedicate to the error correction problem in their way suitable to their domain but such isolated perspectives fall short to explain how systems could develop error correcting features in diversified communications on their own. Therefore, current state of published research still leaves room for unification approaches and for design theories which can help to build systems with the ability to develop communications, error correction and synergetic functionality (i.e. "self-integration").

In figure 6 we see a compact interpretation of the findings in research and what they suggest to me to be the path of integration:

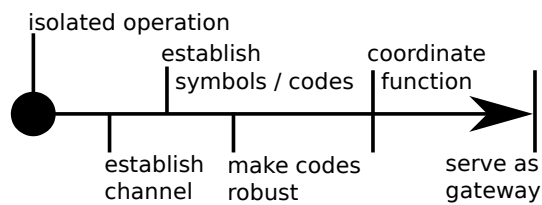

*Figure 6: An interpretation of literature findings how systems will integrate from individual components to coordinated networked components with dedicated behaviors to serve as communication protocols.*

At first systems will have no communication (but no "no effect" on their environment). Certain types of objects can utilize their observations in order to entangle some of their attributes with other objects (e.g by resonance). This yields channels but yet without symbols or codes to be transmitted. Objects must find out what to communicate about and how, particularly how to avoid confusion between "normal behaviors" and "communicative behaviors". Once this is accomplished and interactions become "valuable" to the objects, then those communications must be made robust against noise. This process is continuous and is to be accompanied by a gain in coordination which can be understood as a process of eliminating systematic errors in object interactions.

All this can culminate in very flexible and robust communications which are distinct from other possible behaviors of the involved objects. I hypothesize that this ability to act as a communication multiplexer can become very valuable to third parties with less developed communication capabilities. A shared use of the communication object by the simpler ones should lead to more complex protocols and nesting of protocols making the original objects develop into gateways and communication backbones in the self-integrating ecosystem.

## 3   Digital, Analog and Complex Communication Channels

The basic schema of a communication channel is shown in figure 3: A signal is encoded, transmitted, corrupted, decoded, corrected and then consumed. The flow is unidirectional. A more complete version of this schema is shown in figure 7: The communication path includes various types of encoders and decoders with compression capabilities as were presented by Moon [4].

The communication channel can be described in terms of bandwidth (analog) or more generally *capacity* (volume of technical bits which can be transported per unit of time) but also delay or reliability. More technical properties are the ability to buffer information, type of encoding/modulation or the set of senders/receivers addressable by that channel. Information transported over a channel has meta-properties regarding the communication channel such as real-time requirements. An information can be transported over a channel if the channel can satisfy the transport requirements.

Please note that the channel concept in figure 3 is not just a simpler concept than the one in figure 7: Figure 3 presents an abstract channel and figure 7 presents a digital channel. In most applications a digital channel transmits much less net worth information (e.g. 10 bit/s) than its declared capacity (e.g. 1 Gbit/s) would allow. It will depend on the sophistication of the communication parties whether in any given situation the models in figure 3 and figure 7 can be considered to be at same level of abstraction and hence whether technical and abstract channel capacities are to be considered to be the same.

The model in figure 3 does not make any assumptions about the technology used to constitute the channel but we are used to assume some kind of electronic transmission channels. However, more traditional communications using for example paper can also rely on error correction. In fact, commuting any type of naturalistic objects with states can be generally seen as an information exchange process [41] which can be made robust by applying error correction techniques (cf. fig. 8, top). This sounds like a stretch at first but imagine the following:

Information is usually received in order to define and enable a certain way of processing on a remote object, e.g. a state change. If we send not only a single photon (to do the alteration) but many photons containing a program code which is then used to enable a new processing by the receiving remote object then this would be no different than sending a physical object, let's say an excavator to a construction site in order to enable a certain processing of a remote product (digging holes in the ground).

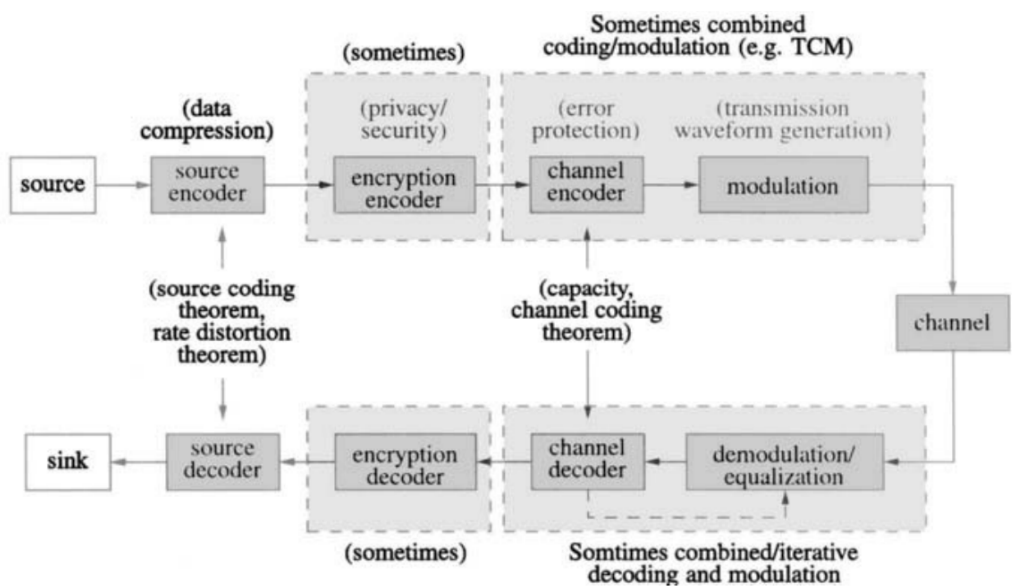

*Figure 7: A more complete version of the model shown in figure 3, as was reported in [4].*

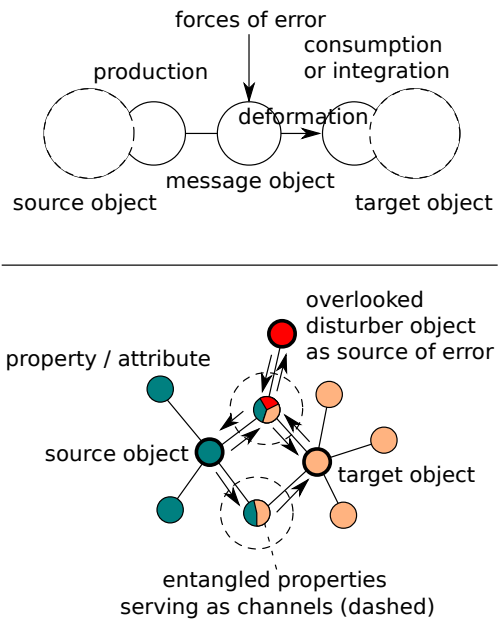

*Figure 8: Two alternative models of communication channels to which are pronouncing the fact of transport and interaction between source and target object above*

Please note that a photon and an excavator are an energy source and information in one item and I do not know of a single example where transmission of information occurs without simultaneous transmission of energy to manufacture intended state change on the receiver, however small. For example, data on a CD looks like an "energy free" solution but sending data on a CD will require to burn it, to transport it mechanically to the new place and to power the laser for reading. Information transport and use of energy are intimately intertwined. This has a fundamental impact on our concepts about communication channels and how we understand their directivity because unidirectional energy exchange is a rare (and maybe only an artificial) phenomenon – usually senders and receivers of energy affect each other to greater or lesser degree.

Anything that we can transmit can and must perform a function on the remote side in order to convey the information. A photon can and must trigger a state change on the receiver side and it will be consumed in this process as much or even more than the excavator which will be consumed in the process of digging holes. If this is true then sending excavators could require error correction. In our example this could be an instruction manual and a repair kit. These objects hold a redundant function or information of the net object which is transmitted and can be used in order to detect defects and to fix them before use.

The source-message-target-object paradigm from figure 8/top highlights the time delay aspect of communications and that units of communications can be arbitrarily more complex than for example *bits*.

Another way to see communication channels is to see them as entanglement of spaces and each object with attributes spans a state space for this object type. This is particularly useful view if the delay on the line is negligible. Two objects mean two different spaces and interaction between the two objects does not occur based on the method-call-paradigm but by influence through space entanglement. Arriving at new states in one space causes changes to points in an entangled space. Not all subspace areas are legal. Which part of space is legal or illegal is defined by constraints in the object model. Objects monitor their attributes and attempt to react to changes of their attributes in the sense as to bring them in consistency with the object model. This can lead to complex synchronization interactions between the objects.

If the object adapts other of its properties in order to be consistent with the externally modified property then we would speak of "information processing" in that object. If the object rejects the change (reverts the state) then we would not speak of "information processing" despite that both processing phenomena are footed in the same mechanism of trying to maintain object attributes consistent with the object model.

The *error source* in this model is then to be understood as a *disturber object* which is entangled with source and target objects but which existence and influence has been not modeled because it was not expected by the communication channel designer. The resulting interactions can entail everything from systematic to (apparently) random error.

# 4 General Concepts

## 4.1 Sources of Error Correction Information

In introductory courses the Hamming 7/4 code is shown as an overlapping of three sets. The syndromes of the three sets (anomalies of XOR-ing the bits in the sets) can be used to reverse conclude which bit has been damaged. For this to work, the Hamming 7/4 code (and many others) transmit additional information (3 bits of redundant information) which can be used to restore original information. In fact, recoding bit streams has the purpose to "uniformly dilute" the amount of information stored per bit: After recoding each technical bit is holding less than 1 bit of information on average ( r < 1 bit ). The theoretical amount of compensable information loss for $x$ net bits is: $d := ( x / r ) - x$, where $d$ is the number of compensable information bits. It means that for example, if we transmit $100$[4] bits of white noise and dilute them in 120 bits of data then we can hypothetically loose 20 bits of data (d) before original sequence becomes unrecoverable.

Unfortunately, the $d$ is neither equal to the extra bits transmitted with the code nor is the representation of the information used to correct the received bit streams. The reasons for this vary on the mathematical properties of the selected code and protocols. Some code streams may even transmit no extra corrective information and yet have enough redundancy in order to correct them. This is because even the original stream can have less than 1 bit of information per transmitted bit on average.

The size of $d$ is for right now a purely hypothetical quantity. All that needs to be agreed upon is that it exists and that it must be provided either by the sender or by the receiver. More precisely there are $n$ nested senders and $n$ nested receivers which are providing redundant information for error correction.

So said, figure 9 is indicating that communication and error correction is not only about correcting elementary entities such as binary codes but any more sophisticated entities, such a graphs, sentences or even complex physical objects. Detecting or correcting received information is relying on other pieces of information[5] which are informationally entangled with the transmitted net information (because totally unrelated redundant information is not usable for error correction). In this context, additional facilities required for correcting transmitted entities are equally diverse as the actual means of communication.

Since the additional information must be "entangled" with the actual "message", it is highly redundant. It can be seen as an overhead which must be transported. I would like to call this a *corrective overhead* as it better suits the broad range of information transport means than the term redundancy. In situations where ontological chunks are transmitted (as would be the case with OWL) I even prefer to speak of *semantic overhead*.

The corrective overhead or corrective redundancy can be supplied either by the sender or the receiver. The receiver can start with minimalistic assumptions of the data such as the choice and parameters of the error correcting code and collect more information over the time in order to increase its error correcting strength (left side in figure 10). The context pool is never empty as even static error correction (where all redundancy is provided by the sender) will require certain assumptions how the code is built and how it is to be exploited (right side in figure 10). In conventional protocols the pool is often implicit as all the knowledge is part of the error correcting algorithm. In other situations the contextual knowedge pool is provided explicitly as is true for example for spelling, grammar and language error detection [33], [42].

In figure 9 there are two examples (*n*=2) showing different content types in information flows: On the bottom the technical bits are displayed. These technical pieces of infor-

---

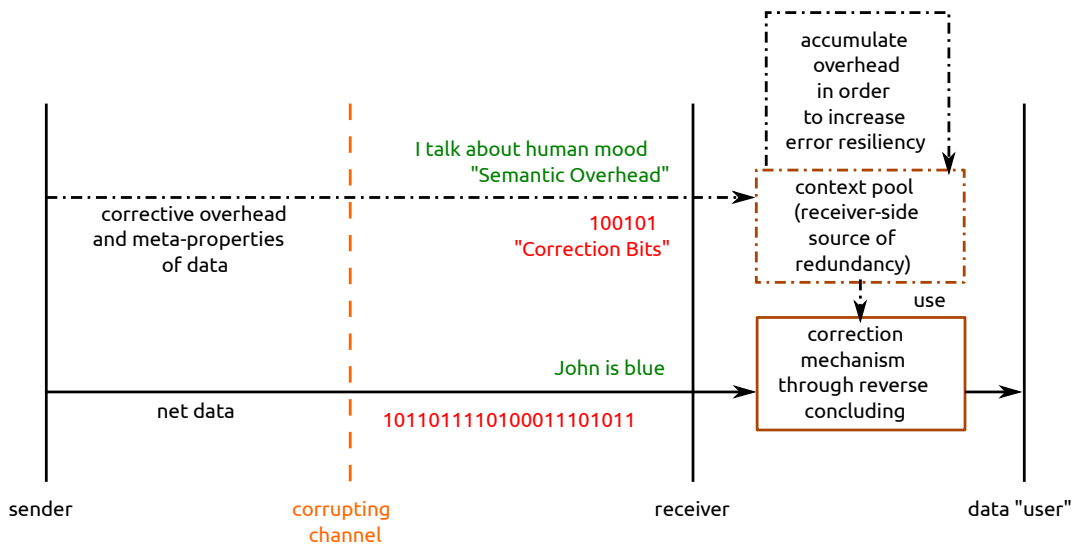

*Figure 9: Redundant information for error detection and error correction can be provided either by the sender along with the message or by the receiver from a communication context. The receiver can exploit natural redundancy in the data and/or build a context from data alone which can then be used in order to provide additional information for error detection or error correction purposes.*

mation require even more bits to be present in order to verify the correctness or intended content (after error correction). Meaning of a bit is defined purely by its position in the sequence. However, the information to be conveyed can be actually of higher level abstraction. The above line of transmission shows the message "John is blue". From the point of view of a higher level receiver this information is erroneous because if taken literary it would imply a blue colored man which is not the message.

Figure 9 emphasizes the importance of the internal context storage ("context pool") as a source of information usable to detect or correct errors in incoming signals / messages. This pool is not of static size. At the beginning of new communication sessions this pool can be almost empty. Sender and receiver must communicate very precisely and the receiver will often rely on requests for further information in order to reconstruct original message of the sender until communication partners accumulate enough information in the context pool. In that initial phase communications should occur over very low noise channels and with high amount of error correcting redundancy attached to it. Later, the amount of redundancy can be lowered while attaining higher levels of error correction (cf. figure 10).

This process is quite natural to us humans: Humans provide this additional information by body language, idioms, formal speaking formulas and the like. Once the communicating team has accumulated enough context information, teams reduce amount of redundancy, speak less formally, issue maybe short sentences, etc. This is the "wordless understanding" in which the amount of technical information is approximating the amount of real net information to be transferred. In fact, if the error correcting information has proven very effective, even less technical information can be conveyed than is theoretically necessary to describe the content of the message. The error correcting mechanisms will simply restore the most likely message until the cooperation on that ground breaks and parties return to richer communications.

The same is achievable with technical systems. In order to attain strong error correction, mechanisms should exploit context pools located at all communication layers. Why is that? The greater the amount of models and their complexity the more symbols and longer symbol sequences[6] are needed which must be protected against corruption. First and foremost, long sequences of data consume a great deal of time or bandwidth to be transmitted. Therefore, it is technically very wise to limit necessary sequences and number of involved symbols to bare minimum in a certain communication frame which is spanning not only a limited number of communication parties but also a particular time frame. In the end, if the communication takes longer than what the communication is about then there is no tactical use in engaging in communications, isn't it?

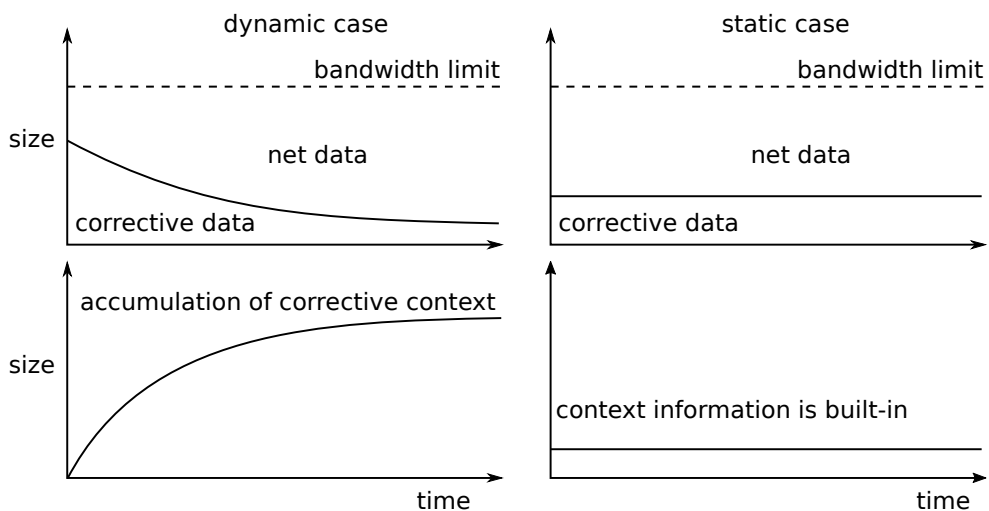

*Figure 10: Redundant information needs not be provided at constant rate and in fact can (even must) be provided a priori.*

Since most intelligent systems permanently learn new things, their symbol and grammar structures permanently evolve. If we wanted to error-correct them in communications we would end up defining new error correction matrices every time when a system's model expands. Worse, this update has more or less to be distributed in the communication frame's parties.

Please note that what I am talking about here is not the same as serializing complex knowledge to binary and then applying one of the known error correction algorithms. A higher level error cannot be corrected this way. Systems must be able to error correct the complete spectrum of symbols (or concepts) they can transmit. We can understand this by analogy: Nesting the involved protocols nests the protocols' states (in sense of protocols as automatons) into flat automatons (or flat protocols). The number of states is exploding dramatically. The job of a flat context pool is to monitor these states and historical transitions between them. Error corrective function would then lie in estimating the nearest legal state after the most likely transitions. For any non-toy system handling such corrections is not practically feasible. Therefore systems must factor out fragments of the state space. You find them in each of the protocol layer but the resulting layers contain more than one possible protocol and hence more than one error correction mechanism. Selecting between these mechanisms can be achieved by using information from higher ranking context pool as shown in figure 11.

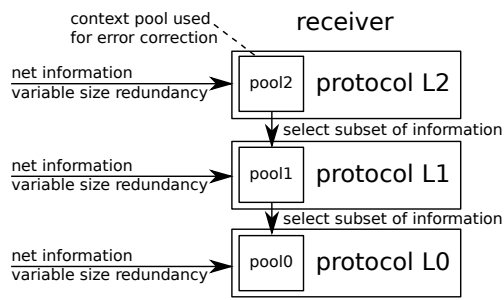

*Figure 11: Cascade of context pools which can serve as "filter banks" [36].*

---

6   Bit sequences are a special case of symbol sequences

Figure 11 shows how nested channels can provide context to lower level pools. Since there is some temporal relationship between symbols transported at all layers (lower streams encode and transport content of higher level streams) we can assume at least a probabilistic dependency between what has been received on higher levels and what will be received on lower levels. If B follows A and A was received and B is encoded 101 then the next expected bit is 1 (unless a 1 has been received, then we expect a 0).

This could be combined with a contextual recoding of symbols. If sender and receiver can agree that the context demands that not all possible symbols but only a particular subset can be expected then error correction can be based on biased algorithms (such as probabilistic ones, e.g. Bayesian decision trees) for error correction. In the next chapter the bias of error correction mechanisms is discussed in more breadth.

If senders and receivers can recode each subordinate layer to accept encodings and correction only to the expected necessary level then what is possible from this is the reduction of communications overhead to a bare minimum. This results in short messages with a minimum subset of symbols for which error correction is easy to define. In extreme cases, a single technical bit of information at protocol in Level 0 needs to be transferred in order to provide a single bit of information to the Level 2 receiver (100% efficiency!). This makes information exchanges very efficient but also susceptible to subtle context changes which have significant effect on the performance of the joint error correction mechanism.

## 4.2 Error Correction, Power Consumption and Self-Calibrating Systems

It is easy to oversee that error correction is not just means to establish reliable communications but also that error correction plays a significant role in extending the radius of effectively practical communications and to reduce power consumption or communications cost [43]–[46]. As a rule of thumb, modern error correction mechanisms raise the range of communications by a factor of roughly 10 if not more, despite that error correction is reducing the communication channel's net capacity by a noticeable fraction. Most literature does not compare power efficiency or cost between protected and unprotected channels, hence I put a qualitative visualization of utility of error correction mechanisms into figure 12.

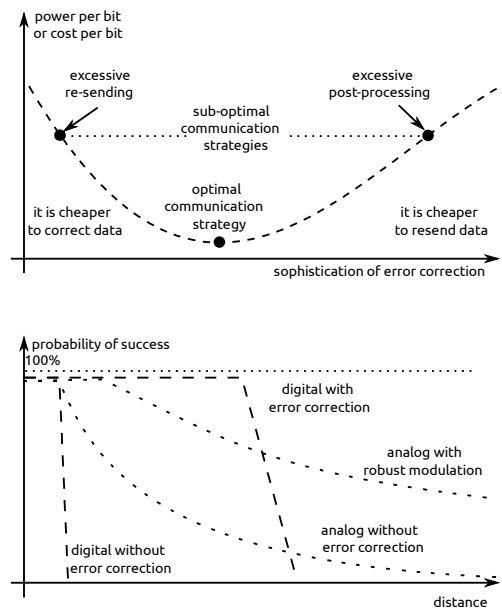

*Figure 12: Utility of error correction mechanism in practical applications.*

Figure 13 shows the role of error correction inside and outside of systems. Internal communications are designed with enough reserves and sufficient signal to noise ratio in order to limit random errors to a negligible probability. However, in recent years this approach was questioned because engineers are trying to find new ways to save power consumption. For example, the concept of self-stabilization in microprocessors [9] is intentionally targeting the optimal communication strategy (cf. fig. 12 (top)): Voltage is diminished or frequency is increased until an optimal amount of correctable computation error is measured [47][48]. Depending on age and environmental conditions, frequency or voltage can be adjusted in order to maintain a sufficient reliability. However, to my knowledge this approach has not been expanded to cover whole systems, including software and hardware, and it has not been exploited to improve overall systems integration. Nevertheless, this opportunity exists.

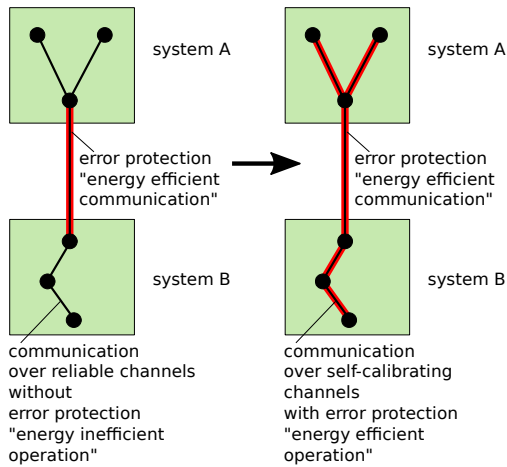

*Figure 13: Energetic efficiency of systems can be increased by deliberately combining error correction with controlled internal unreliability which is influenced over power-related variables.*

In embedded systems with safety requirements, software often consists of several monitoring and diagnosing layers which in conjunction with special hardware such as watchdogs is capable to monitor health of computations. If interaction was allowed between software and hardware about voltages or frequencies, more elaborate interactions in terms of power savings could be possible. However, since digital code is very brittle, controlling CPU power over software could prove very difficult. However, checking reliability of memory is a conventional technique (Double Inverted Storage) which could be used to adjust memory settings. Software could feasibly monitor memory health and adjust memory timings or voltages as in [49].

This view still assumes fully integrated communications which are affected only by random errors, not systematic ones. Systematic errors can only be observed if the system is attempting to abbreviate communication paths or to establish new ones like Cognitive Radio appliances [50], [51] attempt it to do. This would introduce new, ineffective communications between nodes which are not yet connected via one of the links shown in figure 13. However, new links can introduce various kinds of errors on existing links. Therefore, armoring all internal and external communications with sophisticated error correction mechanisms seem necessary to me in order to achieve the goal of slowly self-integrating appliances. Nevertheless, with current digital protocols it is very difficult to move a communication path just a single node as the amalgamate of vertically standardized protocols is too difficult to overcome. There seem to be two options to this: either have nodes support broad range of stacks or nodes can develop on-demand error-robust protocols.

### 4.3 Encoding and Modulation

For the purposes of this paper I will consider encoding and modulation as the same concept (as aspects of a "space-time code" [12]). The term "encoding" will be used when I mean representing information over space and "modulation" when information is meant to be represented development of properties over time.

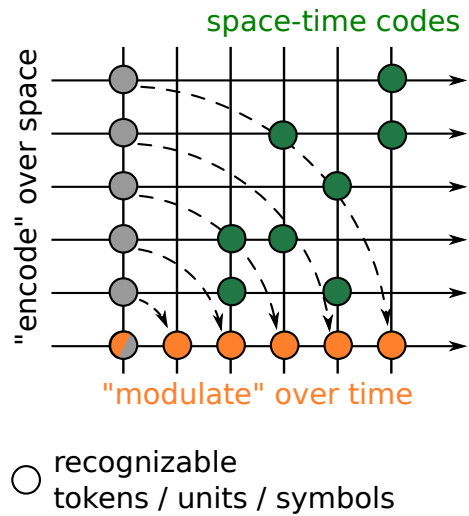

*Figure 14: Modulation, encoding and space-time codes. Each dot represents an emitted symbol, token or other recognizable unit.*

According to figure 14, encoding is essentially the same thing as "modulation" but comes from the analogue communications domain where information is imposed onto electro-magnetic waves, waves which are "modulated". However, from a systemic or information scientific point of view, photons are just some objects with some modifiable but unreliable attributes (and hence susceptible to random error) which can be used to encode more or less (today binary) information.

### 4.4 Adaptive Error Correction

Other opportunity to optimize the protocol for detected error is to change the encoding or "modulation" (as e.g. in [52]) .

Since senders and receivers can agree to use different rules to encode data in order to isolate the disturbing party and since the disturbing party could act very much differently than a uniform or bulk/burst disturber, the question comes up if they can isolate noise sources more selectively.

In general yes as there exist examples of this: [53], [54], [55]. Adaptivity relies on a minimum of feedback given from receiver to the sender as is shown in figure 15. The receiver reports how many errors it receives and whether repair efforts are successful.

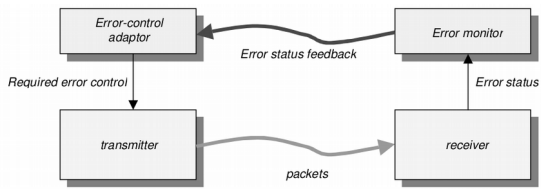

*Figure 15: Principle of adaptive error correction based on forward and backward communications (copied from [56])*

How can we imagine adaptive error correction? For example, disturbers could be triggered by a particular bit pattern and disturb a particular value more frequently. In that case receivers and senders could recode particular values to be more robust. This could occur very similarly to the Huffman coding. However, Huffman coding is for compression and chooses longer representations for less frequent symbols. When applied to error correction, codes get the longer the stronger they are observed to be disturbed.

The above described feature is more often used with analogue communications where certain properties of electronics lead to an uneven distribution of values which are otherwise uniform. For example a sender could distribute the value 1,2,3… on frequencies 1, 2, 3 kHz and so on but doesn't do so because lower frequencies are for some reason more difficult to filter apart (due to some electronic idiosyncrasies). In that case values could be spaced apart unevenly: 1, 5, 7, 9, 10, 10,5, … kHz. This way disturbing input at low fre-quencies doesn't confuse the receiver unit to misinterpret a 1 for a 2. This is the "analogue way" to recode symbols in order to improve a signal's error correction capability. In this hypothetical case, dynamic error modeling could mean that senders and receivers of analog signals adapt, let's say, by using a gamma-factor in order to control the receiving function's non-linearity.

Yet another possibility to correct random errors is to reallocate the mechanism to another layer. In mobile communications senders and receivers can agree to send data over different channels or at different times. The superordinate layer responsible for reassembly of original data would then correct messages / signals in a way that was not possible by the particular subordinate sender-receiver-pairs.

## 5 Sources of Errors

### 5.1 Random Alterations

For sake of simplicity, errors are often modeled as more or less random events. Such random errors are governed by statistical distributions. Error correction mechanisms can be optimized to the expected distribution. For bit streams, commonly, correcting mechanisms concentrate on sporadically distributed or bulk/burst bit errors. Implementing errors for the wrong distribution are either inefficient or do not work. Therefore, senders and receivers are characterized by their *expectation* what kind of error they should deal with. Such expectation can be either statically built-in (by engineer's choice) or can be more or less developed by higher level monitoring services.

In analog communications, the sender and receiver can adapt their encoding and the amount of transmitted corrective information in order to better deal with encountered errors. For this very purpose RX/TX units usually implement various legacy protocols which are often more robust against errors. They are often also slower in terms of used technical frequency which helps to detune from the noise source (the error source). Since the purpose of changing frequencies is to detune from the noise source, the speed of communication can also be raised. Whether this is possible will depend on the nature of the noise source. If the noise source would become more intense (as is often true for analog transmissions) with higher frequency then communication speed can only be lowered.

In contrast to analog technology, digital technology is used to transport heavily nested protocols. Despite that bit errors are the first thing to think of when discussing randomized error sources, indeed any protocol layer can be affected by random errors even if the underlying protocols did work without error.

For example, an XML serializer could have an implementation error and rely in a certain part of its code on a zero-value initialization of a variable. Because the pointer used to access the variable changes during the execution to point to different places, it could hit a place in memory which is not zero (because it has been used by a different program part before).

What could be the resulting error? For example, the closing tag could be missing a slash:

```
<taga> content <taga>
<tagb> content </tagb>
```

Clearly, given some additional analysis, the XML deserializer could detect that the second taga is not nesting a new statement but closing the first one and then correct it.

Despite that the programming error is systematic, the effective errors are sporadic. It will be the XML deserializer (decoder) which will have to deal with incorrect XML data because the error was introduced by the XML serializer (encoder). Any error protection mechanism applied by a lower level bit streamer is not going to catch the error.

### 5.2    Omission Errors

A special class of (random) error is the omission error. The below XML example is actually an example of this kind of error because the closing tag for taga is missing.

```
<taga> content </taga>
<tagb> content </tagb>
```

Clearly, given some additional analysis, the deserializer could detect that taga is not nesting tagb and close it presumptively. Interestingly, detecting omissions on symbol-rich protocols seems to be generally easier than on binary sequences. Detecting omission with bit streams is more difficult which is why technologies attempt to send a bit no matter if the bit has been retrieved correctly or not. The order and grammar of bit sequences then defines the meaning of the bit.

Detection of omission errors can be attained in several ways. The sender could for example announce the length of a message a priori or a posteriori in order to help the receiver detect if the data is suffering from omissions (such as missing message tails). For long messages, the sender could decide to split a message into chunks and enumerate them. This allows the receiver to retro-organize the data fragments and request for missing chunks. If the sender does not know the length of the message a priori himself then it can employ a chunk chaining technique. For example, for each chunk a CRC value is computed and is then multiplied with the CRC of the previous chunk. The receiver will only be able to compute the transmitted multiply-CRCs correctly, if the data has arrived completely and in correct order.

Protocols used to communicate structural information can be designed in such a way that symbols representing certain structural features are occurring more than once in a stream. This redundant transmission of structural information can be used by the receiver to detect various kinds of errors – including omissions. In figure 16 an error is detected because the green node occurs in two different positions. Albeit, this is not automatically holding a clue how to correct the error.

There exist techniques which can fix structural errors by reconstructing a set of possible nearby correct structures and chooses as explanation the structure with smallest amount of diversion from mutilated structure (minimum edit distance). This technique has been most widely explored for compilers [57]–[59] as compilers are notoriously bad at giving hints what and where the true error is and do not automatically provide the ability to make correction propositions.

Dealing with omissions is requiring completely different elements in error correction than is usually addressed with error correcting

& error detecting codes. It requires an information selection capability and a selection confirmation mechanism which is usually not not part of any encoder / decoder. Moreover, utility of such mechanisms is sometimes low if a reverse communication channel does not exist because often the only way to fix an omission is to request anew the missing parts.

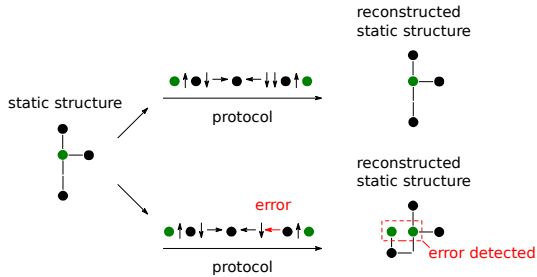

*Figure 16: A non-binary example of a protocol and how it relates to a static data structure. Sending the green circle-symbol is helping to detect certain types of error.*

### 5.3 Erasure Errors

Erasures are very similar to omissions and we know several binary codes which are known to handle erasures, such as LDPC or Reed Solomon codes. The latter are used in QR-codes (fig. 17). A difference between an omission and erasure is that in case of an erasure position and scope of corrupted data is known. Whatever has been received during an erasure yields no information to the error correcting decoder.

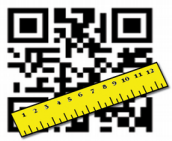

*Figure 17: Erasure on a QR-code*

The difference between erasures and omissions is shown in figure 18. The omission has caused the transmission sequence to be shorter. After an omission it is not known which section has been omitted. On the contrary, an erasure has affected the same region in figure 18 but the error correcting mechanism can at least know the size and position of the error. This information is usable in order to recover the original information in some cases: In figure 18 positions 7..9 are "occluded". No redundant information for er-ror correction was transmitted. However, a knowledge-intensive protocol decoder could know (or deduce) that these sections contain expected values in a particular communication context.

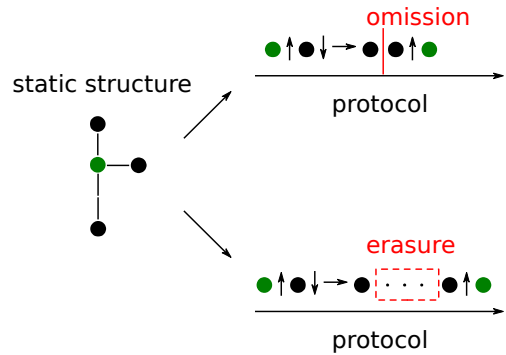

*Figure 18: Omissions and erasures*

### 5.4 Systematic Errors

An important class of errors is the class of *systematic errors*. The simplest type of error imaginable as a systematic error is the *signal offset*. Each message is in whole consisting of a single numeric value which is false.

If the offset is static (and hence systematic), any value transmitted ends up received with a certain value above or below the real value. The receiver can correct offsets in input by simply subtracting the offset from it.

Since offsets are such a frequent source of systematic error, many systems implement correction of offsets in one way or the other. Systems may even monitor their development and adapt their offset-prediction every now and then. In signal processing this process can be understood as high pass filtering. In some cases Kalman filters can be used to measure and filter out systematic errors in signals [60].

However, many more examples of systematic errors can be named. A bad mapping would be one such example. Bad mappings have a structural and temporal effect, depending on the reason for transmitting the information. In figure 19 a temporal effect is demonstrated: The driver system is a spontaneously operating automaton. It sends it state to a partner system (the driven system) which is also a spontaneously acting automaton. If the driver system is a little bit faster or slower than the driven system, it will push its state once in a while a state forward or a state backward. For this to occur correctly the communication channel must faithfully transmit the driver system's states (a,b,c,d).

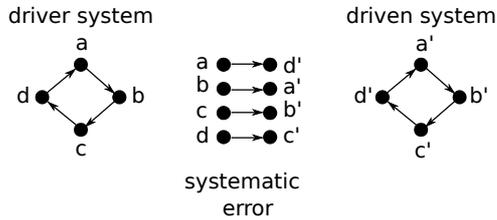

*Figure 19: Example of a systematic error: Bad papping of symbols between the input and output space leads to a systematic phase shift between driver system and driven system.*

However, figure 19 is showing a systematic mapping error between the input and the output symbols. The consequence of this mapping-error is that the driven system is forced to run in a phase shift.

Correcting this kind of error requires a reverse mapping function which is modeling the original error. If the communication channel cannot be corrected (in order to remove the bad mapping) and, ideally, if no information is lost then an "adapter mapping" can be used to correct the transmission error.

The inverse mapping is equally necessary if the receiver is not affected temporally. This is true, for example, for any quantitative distortion. Creating complex reverse mappings (inverse functions) used to correct systematic errors is if often called *calibration*. A calibrated system is correcting systematic errors in a more or less continuous input space. If the mappings are more complex than some stretches, swirls and rotations, it is usually called just a re-mapping.

Combining corrective mappings related to static and temporal effects is usually described as *adapting*. Very complex adapters with internal reconstruction of higher order semantics are usually called *gateways*. In sense of error correction, a gateway is nothing more than a fancy, prepended mechanism for correcting complex, systematic errors of the communication channel.

## 5.5 Ambiguity and Errors

Usually, handling and resolving ambiguity is considered an artificial intelligence or usability topic but in terms of communications it is an error correction topic. Depending on preference, it translates into an omission or erasure error. For example, if the problem is understood as omission then the system should detect ambiguity and request additional pieces of information in order to resolve the ambiguity. More technically speaking, the amount of bits provided for a particular symbol "blue" was not a complete sequence. Expected were "blue:as_colored" or "blue:as_mood". After requesting the "correcting" pieces of information from the sender, the sender will complete the transmission by sending either ":as_colored" or ":as_mood".

If the problem is seen as an erasure then the perception of the receiver is "blue:____". The receiver will now try to restore the blank section with pieces of information it already has or can acquire otherwise. The two views are equivalent to the effect.

# 6 Error Detection and Correction

## 6.1 Problem-biased Methods

The term *bias* is used here to describe a family of error correcting methods which are offering different levels of protection either to symbols or to sections of data [61]. They will show bias toward protecting critical parts with higher priority and risk uncorrectable errors in less critical areas of data.

Most error correcting codes which are developed for technical communications are without bias. That is they will correct the first bit no worse than the third or any other. They must be indiscriminatory as they do not know what kind of content they will have to transport. Applications relying on indiscriminatory methods must choose a code according to the most critical piece of information that must be reconstructed. The other parts of data are "overprotected" - one could say. This yields a bad balance between cost (energy used to process stream or additional time necessary for transmitting additional redundant bits, etc.) and utility must be improved.

One way do this is to use codes with a variable error correcting strength. Such codes can be adjusted dynamically in order to transmit bits of highest criticality over bit positions which have the best data-to-noise ratio. In figure 20 we can see four kinds of bits in a message:

a) unconditionally reliable bits (very difficult to corrupt or cannot be corrupted)

b) conditionally reliable bits (these bits are reliable for particular conditions)

c) conditionally unreliable bits (these bits are unreliable in particular conditions)

d) unconditionally unreliable bits (bits convey almost no information).

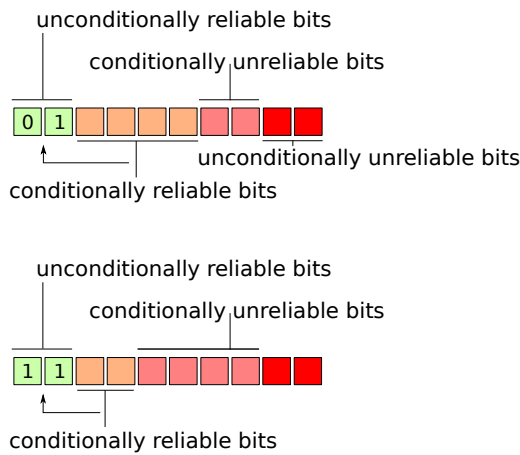

*Figure 20: Usable bits of a code may vary depending on the characteristics of the bit-source.*

Distribution of correctable codes given any particular model of reliable and unreliable bits can be then interpreted as a noise robust spacing of codeword representatives in a space like shown in figure 21:

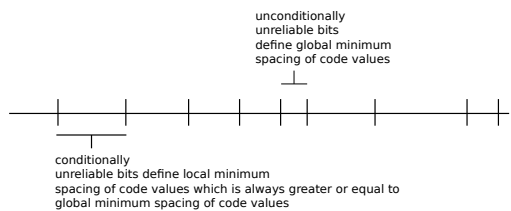

*Figure 21: The effect of bit noise on spacing of symbol representatives. The greater the spacing the easier to eliminate transmission errors.*

## 6.2 Pattern Classification as Means to Correct Errors

The concept of distributing codewords in space can be expanded to more dimensions: Figure 22 shows how codes can be unevenly distributed in spaces giving us a "biased" code set in two dimensions. This way each code is addressing an associated error model. The organization of space representing a particular class of objects (the "code") allows more or less error in particular directions. Deciding whether an observation is matching a particular class of objects (which code must be attached to an observation) is a classic pattern recognition task. Observations in pattern recognition suffer from a lot of noise and the task of pattern recognition is to remove resulting variability from the labeled output.

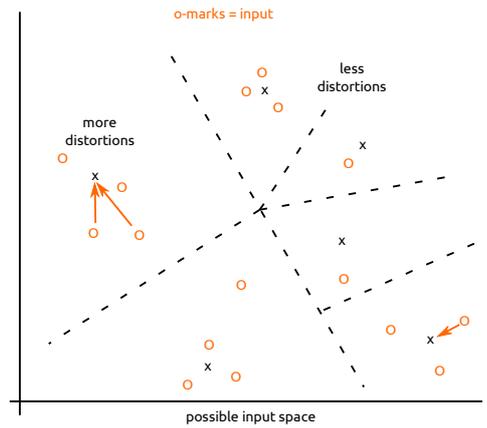

*Figure 22: Error correction in a non-digital input space based on nearest-neighbor classification (adapted from [6])*

How could this work in practice? Let us assume that there are two signal lines of input for the receiver. If the signal lines were used to transmit only binary information (encoding points are combinations of LOW and HIGH) then only two bits of information can be sampled at each sampling round by the receiver. However, the amount of information transported on the same wires could be increased by 50% by choosing other encoding points[7]. Such working points are specific to the technology in use. In order to transmit 3 bits per clock we must discern between eight states on the two wires (figure 22 shows only six for clarity) which are marked with an x.

---

7 In communication applications these two "wires" are commonly frequency and phase.

The communication works as follows: The sender attempts to set a certain voltage-pair on the two wires and chooses one of eight prototypes for this depending on which of the bit sequences it wants to transmit. Since the the voltage levels are an analog & physical signal they are susceptible to various kinds of noise effects but the noise can be different for each of the prototypes. This can be taken into account by optimally distributing the eight prototypes across the space.

Once the prototypes have been distributed and defined, the receiver has then the task to record the voltage level on both wires simultaneously and then to decide for them jointly which digital information was conveyed with the recording. In order to achieve this, it must segment the input space measured in Volts and use this segmentation in order to classify the voltage values as one of the eight three bit sequences. As long as the  noise on the line obeys certain limits, the classifier is performing an error correction function. It can detect an error (a deviation from the template) by comparing the recorded vectors with the prototypes and report a fatal correction error if the measured vectors are too far from the prototypes – this would be the function of the nearest neighbor classifier.

Most physical modulations used for communication are using some sort of classification based on amplitude, frequency or phase shifts. Receiver chips are basically nothing else than hardware-classifiers [11].

The classifier approaches works best if the observations made by the receiver are close to the true signal. In figure 22 the orange circles (samples) are close to the actual prototypes.

### 6.3   Error Correction Methods for Digital Streams

The classification occurring in figure 22 gets the more difficult the closer the samples approach the boundaries of the classification space. Unfortunately, exactly this happens with digital codes which are favored for communications because of the good detectability of states in presence of strong noise. In fact in binary messages samples lie exactly on boundaries as is shown in figure 23.

There are many possible errors in such scenarios yielding exactly the same Euclidean distance to particular code reference points. This symmetry can be only broken up if symbols are distributed over additional (redundant) data dimensions. Therefore, in digital communications error correction using binary representations requires additional space in order to sort out bit errors.

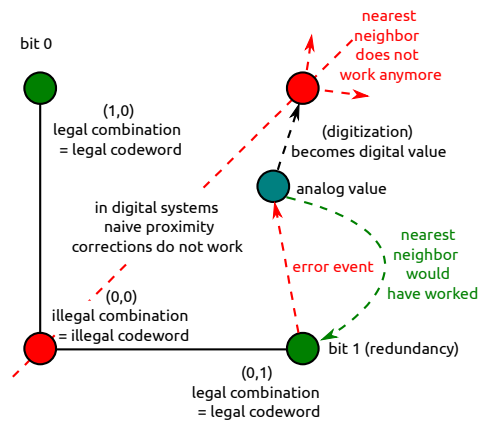

*Figure 23: In binary spaces erroneous codes may not suggest any fix. Those codes can only be used to detect errors.*

This additional dimensions get expressed as additional technical bits which must be transferred over the wire. Trellis codes [62] or lattice codes are known ways to populate the additional space with correctable codes. However, from the point of view of classification nothing changes. If input prototypes are organized along additional dimensions in order to generate a greater proximity of erroneous codewords to most likely legal codewords then a nearest neighbor classifier would do an excellent job to detect and correct the input. It would probably do even better if the input signals were not binarized before processing (and inertia of electronics were exploited in order to increase the distance to undesired prototypes).

In figure 24 all 16 codewords of the Hamming 7/4 code are shown and example of a error corrupted by one bit. Visualization of the codewords can be seen here:

https://en.wikipedia.org/wiki/Hamming(7,4)

| error word | 0 | 1 | 1 | 0 | 0 | 0 | 1 | Distance (Bit) |
|---|---|---|---|---|---|---|---|---|
| legal word 1 | 0 | 0 | 0 | 0 | 0 | 0 | 0 | 3,00 |
| legal word 2 | 1 | 1 | 1 | 0 | 0 | 0 | 0 | 2,00 |
| legal word 3 | 1 | 0 | 0 | 1 | 1 | 0 | 0 | 6,00 |
| legal word 4 | 0 | 1 | 1 | 1 | 1 | 0 | 0 | 3,00 |
| legal word 5 | 0 | 1 | 0 | 1 | 0 | 1 | 0 | 4,00 |
| legal word 6 | 1 | 0 | 1 | 1 | 0 | 1 | 0 | 5,00 |
| legal word 7 | 1 | 1 | 0 | 0 | 1 | 1 | 0 | 5,00 |
| legal word 8 | 0 | 0 | 1 | 0 | 1 | 1 | 0 | 4,00 |
| legal word 9 | 1 | 1 | 0 | 1 | 0 | 0 | 1 | 3,00 |
| legal word 10 | 0 | 0 | 1 | 1 | 0 | 0 | 1 | 2,00 |
| legal word 11 | 0 | 1 | 0 | 0 | 1 | 0 | 1 | 2,00 |
| legal word 12 | 1 | 0 | 1 | 0 | 1 | 0 | 1 | 3,00 |
| legal word 13 | 1 | 0 | 0 | 0 | 0 | 1 | 1 | 4,00 |
| legal word 14 | 0 | 1 | 1 | 0 | 0 | 1 | 1 | 1,00 |
| legal word 15 | 0 | 0 | 0 | 1 | 1 | 1 | 1 | 5,00 |
| legal word 16 | 1 | 1 | 1 | 1 | 1 | 1 | 1 | 4,00 |

*Figure 24: Decoding a Hamming 7/4 code using a nearest neighbor approach. Distance is the Euclidean distance between error code e and any legal code l (||e-l||² or [bit]).*

Get the sheet from here and try it out: *http://lodwich.net/Science/hamming_code.ods*

Resolving a Hamming code in this example is simply done using a nearest neighbor classification in order to demonstrate that classification is a conceptually suitable approach to understand error decoding. However, when the number of codewords and their length start growing then nearest neighbor approaches can become computationally prohibitive.

### 6.4 Contextual Codes

Furthermore, the distribution can be designed in such a way so that transitions between nearby prototypes are improbable. To reduce nearby occurrence of prototypes further we can chose to allow only "far away codes" giving us contextual codes (e.g. [63], [64]). Figure 25 demonstrates such a coding method where the meaning of a code depends on the previous code: The bit triplets a-h are used to encode ones and zeros. Which of the letters is a one and which one is a zero depends on the last code. Errors will naturally cause code shifts in to the left or to the right. Given last code the new codes are chosen far left and far right on the code sphere. This guarantees an almost equal error resilience between current code and the new codes which have become legal.

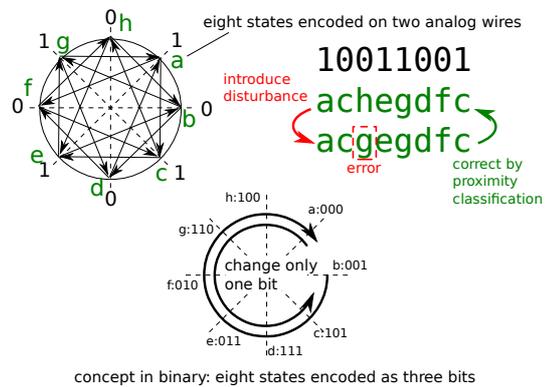

*Figure 25: Contextual encoding of bits in two analog dimensions used to communicate binary data (adapted from [64]). Valid codes depend on previously received codes, hence it is contextual code.*

*Please note that phase-frequency coding is used to code three bits of gross data while in the above example there are only two possible follow-up states. For example, in context of state c, the only legal follow-up states are h and e. The states c, h, e are distributed (roughly) equally on the coding circle. This makes it easy to a) detect that a new bit is transported and b) that likely to occur errors will stay in proximity of the intended bit code. Therefore, the illegal follow-up code g can be corrected into h (h is nearest neighbor to g).*

# 7 Correcting Errors in Cybernetics

*"Information is the difference that makes the difference"*

## 7.1 Relating Errors in Communications and Control

A major field of science, the control theory, concerns itself with error correction. In control systems the problem lies herein that a command sent by the sender is not properly implemented by the receiver. Since control applications intend to attain a difference of zero between the control variable and the controlled variable, we can understand any basic closed-loop control scenario as an attempt to communicate a piece of information to a receiver (the plant) which can corrupt it in such a way that the sender can nevertheless find an alternative representation which is able to survive the erroneous distortion.

Characteristic of error correction performed in control setups is that the correction is not performed by the receiver – it is not a forward error correction but a "backward" error correction. The receiver reports the value it received back to the sender and it is the sender trying to fix the error. In conventional backward error correction the scenario is limited to receiver-defined retransmissions of data. However, I postulate that this operation can be seen as a special case in a range of possible interactions within a general case in which a) the sender relies on a back-link in order to decide what and how to transmit and b) in which it shares correcting intelligence with the receiver.

The control-loop application is simply yet another special case within the range of the general concept in which the sender is having significantly more possibilities to react to an error while the receiver is particularly passive about it. The interaction between senders and receivers can be then understood as the attempt of the two parties to arrive at the same codeword/symbol which temporarily stand in an equivalent semantic relationship until error conditions change and equivalence is destroyed.

## 7.2 The Information Modulation Process

Control applications deal with quantities and rates of development and influence between those quantities. Users of Simulink, Scicos or Modelica like to model quantitative influences as signal flows between processing blocks. Ideally, signals take immediate effect but for many reasons (integrators, latch registers, etc.) this is often not the case. Realistic controls and plant models exhibit various types of delay. System models as shown below can exhibit a propagating wave of excitation.

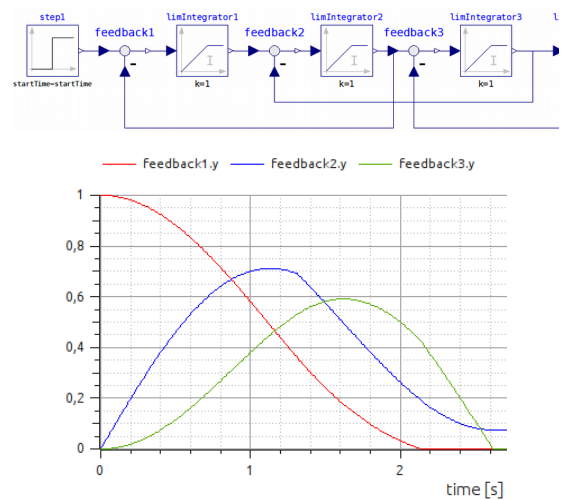

*Figure 26: Activation cascade: feedback 1 activated first, then feedback 2 activated, then feedback3 activated. Finally, all feedbacks are deactivated. The above example dilutes the signal like many physical communication systems do.*

These ripples of excitation are usually avoided by engineers or at least designed in such a way that no self-excitatory feedback is collected by the system. Nevertheless, the propagation of "the difference" is very similar to what one would expect from a propagating single time signal which could hold a piece of information for a signal interpreter.

This idea is reused here in order to explain the process of modulation of messages emitted by a communication system in general because a self-integrating system must also find out when and which type of message to emit in order to perform its role in a system.

For this purpose, figure 27 shows the basic setup for communication explained as a regulatory process. The sender and receiver are connected via a two-edged link. The dashed boxes represent a constant amount of information that can be transmitted per unit of time. The boxes travel from sender to the receiver and back. The *channel capacity* or *channel bandwidth* is the equivalent of the number of boxes multiplied with their sizes multiplied with their traveling speed, in total expressed as boxes consumed per unit of time. Some of the boxes are never filled – in that case they show up as virtual messages. The size of the box is purely theoretical for our purposes here, it can be worth a single bit or a Megabyte.

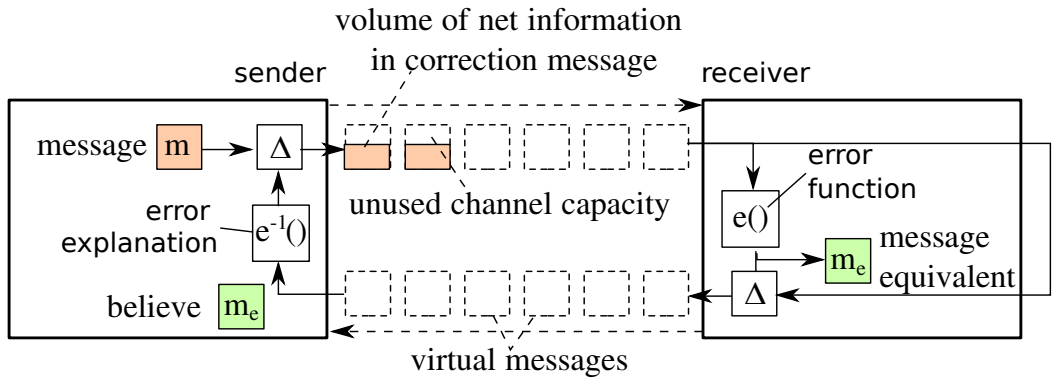

*Figure 27: Sender controls receiver's interpretation of the message in a control loop scenario. The amount of information indeed exchanged is the result of a difference function between message and confirmation.*

I keep empty boxes in the picture because even if they have no content they might still represent something real in transmission like for example zeroed frames which have to be transmitted but will lead to no processing on the receiving side. The sender of the boxes (this can be the sender on the upper link and the receiver on the lower link) can fill the box to any fraction of its capacity.

The $\Delta$-function defines how much information needs to be transmitted. In linear control scenarios the *minus* operator is an approximation of the $\Delta$-function. In strict approaches the controller must reduce the amount of bits used to encode the correcting value. The closer the error to zero the shorter the representation of the control signal until the signal disappears (error = 0 → bits = 0). However, if an unpredictable force changes the parameters of the receiver's plant then the amount of net information transmitted increases temporally.

That means that in closed-loop control scenarios the amount of information transferred from the sender (controller) to the receiver (plant) varies over time despite that the signal my require the same channel capacity for technical convenience, as controllers are usually designed in such a way that their output is sampled in full width at each time - for example it will be sending the same 32bit value at 1kHz all the time.

## 7.3 Information Modulation by Controllers used as Monitor Devices

In order to better understand how information volume is modulated by a sender it is sometimes worthwhile to change the perspective a little bit. In the following example I sketch a controller which is used as a monitoring facility in a microcontroller. In that example, the sender can be able to correct an error reported be the receiver in a single step – for each computation round the system can overwrite any size of memory and the communication channel is not buffering any "virtual message boxes". In this example, the receiving plant is a part of RAM which suffers from occasional overwriting. The sender is a monitoring system which will compute a difference between a store away copy and the particular piece of RAM. Legal writing to RAM is only valid if monitor is disabled, memory is changed, copied to store-away memory and then monitoring is re-enabled. In such example any deviation of RAM from the back up memory will lead to an immediate restoration of RAM. The amount of single-time writing done to RAM memory is the amount of corrective information necessary to correct an error (illegal overwriting). Despite that this example is not obviously a closed-loop control scenario, it indeed is: The sender receives from the plant a vector signal (memory) which is subtracted from the reference vector (back up memory). The resulting delta can be added to the memory in order to restore it in one go. In a naive technical implementation all of memory would be overwritten but a smarter solution will only write the bytes which have been mutilated. Even smarter solutions would only flip the incorrectly flipped bits. Therefore, the amount of error correction information surmounts only to the number of bits which have to be reverted – and this amount can vary from case to case.

### 7.4 The Knowledge in the Model, its Acquisition and Maintenance

Particularly interesting element in figure 27 is the error function $e$ and its inverse function $e^{-1}$. Backward error correction relies on successful model inversion. Nowhere is this better understood as in linear control. In control application the controller is usually designed to invert the signal distortion occurring in the plant.

The error function in figure 27 is nothing else than the distorting behavior of receiving party. If the receiver is adding a 2 to all received values then subtracting a 2 by the sender is necessary in order to have the receiver interpret the value as intended.

The error function can be considered noisy by some amount in order to explain signal deterioration of content in the "message boxes". The range of volatility of $e$ will vary depending on the application. In linear control the systems to be controlled are defined at design time and will alter their characteristics only little over time. Nevertheless, even slow wear out of the plant will require strategies how to align the controller's inverse model with the current true model of the plant.

This process can be understood as the adaptation of $e^{-1}$ to $e^{*-1}$ when $e$ changes to $e^*$. There are applications where changes to $e$ can be communicated using a dedicated channel but in the absence of a separate channel, the sender must use the existing communication channel in order to find out whether $e$ has been updated to $e^*$. This can induce exchange of additional information (signals or messages) which is not explainable by information modulation process as has been explained before. For example the system could be designed to transmit a basic amount of information just to confirm sound relationship between $e^{-1}$ and $e$.

What does it mean in context of conventional digital communications? Well, codes used in analog applications are contiguous and well ordered but in digital communications this is often intentionally not the case. Whatever the encoding order applies, it is defined in a *code set*. If the interaction in the control plant is understood as a communication channel between a sending controller and a receiving plant (quite analog to the bi-directional communication channel discussed in [10]) then the error source can be said to have distorted the code set of the receiver. The sender must now find out which code set to use instead in order to restore correct interpretation of messages and signals by the receiver.

In linear control, no method represents this idea better than the step or Dirac pulse excitation methods. In analogy to this method a communication partner in the role of the sender can perform "crazy Ivan maneuvers" (CIMs) and observe the responses of the receiver. For practical reasons he should not only look at the feedback over the back-channel but on all channels possibly conveying information about the receiver. In fact, CIMs can convey the presence of additional communication channels between senders and receivers which were not known before. This is a crucial function if the goal is to achieve communication channel independence. Once systems discover additional channels of communications they can start to reserve particular uses to them, such as $e$-$e^{-1}$-synchronization. This will make the systems appear coordinating through communication from the outside.

### 7.5 Temporal Characteristics

The ideal controller will keep the time measured between the first report of error and its complete cancellation as short as possible. Since the amount of information considered necessary to correct a particular error is deemed constant, shortening of the total correction time is only possible by transmitting more information per unit of time.

If this process is understood as a re-configuration process then the speed of this process is limited by link speeds $n$, $m$, $q$ and $r$ of the representative minimal omega unit architecture [10] where $n$ is the link between the controller and the plant and where $r$ is the feedback connection. In figure 28 a general landscape of error is depicted as discussed in [10]. In control applications "*transmission errors*", "*constraints errors*" and "*deterioration errors*" are of main focus. Indeed the controller or the input to the controller can also suffer from errors but in control theory this concept is understood as an external problem which needs to be addressed by outer control loops.

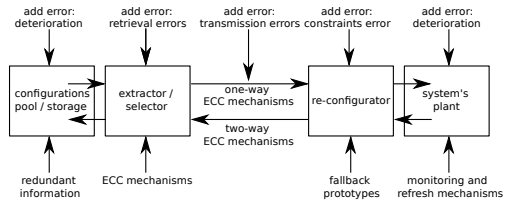

*Figure 28: The communication channel where access to storage is performed by an extractor (the analog of a memory controller)*

The bandwidths $m$ and $q$ and mechanisms used to implement input in plant greatly contribute to the plant's temporal characteristics. This paper goes one step further and assumes that any dynamic behavior is the result of limited bandwidth re-configurational activities in the systems. This is not difficult to conceive, as all physical systems consists of atoms which have to exchange information in order to interact. All systems made of atoms retain that characteristic. This means that the total of bandwidths subsumed in the minimal omega unit bandwidths limit the inflow of information into the plant and hence determine the effective length of time necessary to correct the error.

In figure 29 the dynamics of this process have been depicted qualitatively. Ideal control results in an instantaneous surge of information flow between controller and plant as shown as a spike in figure 29. For realistic control applications the ideal case is impossible to attain. Instead we ask to design optimal controllers. Optimal controllers will need some time for error correction and this time is the result of the ratio between information exchange demand and involved bandwidths. Sub-optimal controllers will implement search policies which will waste available bandwidths/channel capacities and will hence take even more time to correct an error.

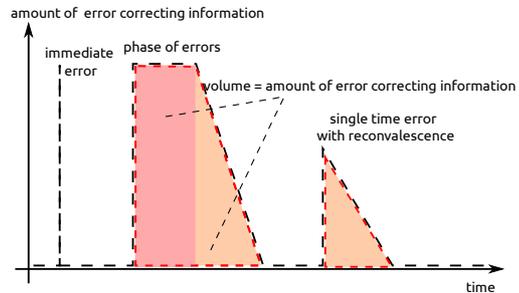

*Figure 29: Dynamic production of error corrective information in a closed-loop control scenario.*

Readers who are interested how this can be understood more formally can find several works on this matter, particularly in context of networked or distributed control systems which would require a long transmission path [65]–[67] as shown in figures 30, 31 and 32.

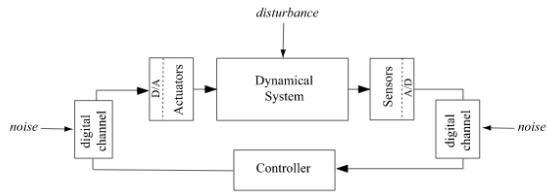

*Figure 30: Role of communication channel in a control application as shown in [65].*

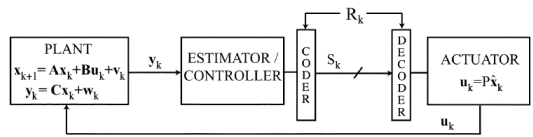

*Figure 31: Alternative arrangement from [66]*

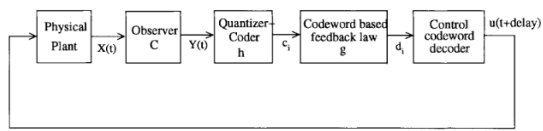

*Figure 32: Alternative arrangement from [67]*

### 7.6 The Generalization in Question

The concept of Δ-functions on sender and receiver side is also an allusion to the initially stated idea that information is the difference that is making a difference. Even if not undisputed[8], the model in figure 27 makes clear what it means: Information must be seen not as any difference that would cause some effect, but instead as the difference in signal content which can be interpreted by the receiver for the purpose of updating its internal state (which is the intended difference that is to be made) – this is the reason why information is so popularly explained in the catchy phrase about the "difference that is making the difference". This is absolutely compatible with the way we are nowadays understanding and measuring information: The change in signal which is toppling the system between two equally likely states is worth one bit of information.

The model of virtual messages which are being exchanged back and forth is also an allusion to Motley's idea that you cannot not communicate [68]. The justification for this model is that systems depicted in figure 27 are often entangled via a communication channel that is not immediately visible to an external observer because it only transports "empty virtual boxes". Those boxes arrive at receivers and senders alike and confirm them that their believes ("input buffers") and error models are corresponding. Any change on either side will result in signals emitted by the parties in order to re-synchronize. This is very similar to electric charge which starts to radiate even if only the "frame of reference" starts moving. Radiation of messages or information can be the result of a change of context between two systems. The Δ-function and the error function of the model in figure 27 is capable to explain this phenomenon.

Motley's figurative statement about inability not to communicate is not undisputed (e.g. [69]). Interestingly, the opposite seems to be also true: Just because systems are exchanging visible and measurable signals with each other, it does not imply that they are indeed communicating with each other. Senders and receivers can exchange signals which represent "empty virtual boxes" but which are visible to the external observer and hence could suggest to him that the two systems (senders and receivers) are communicating.

This is said to avoid a fallacy in which the amount of observed signal in a particular environment is an indicator of "much" or "little" communication, of "many" or "few" communication channels. A quiet environment in which only few things "happen" can constitute a much, much more communicative environment than a loud environment in which there is basically a lot of noise and only little genuine communication (in sense of truly transmitted bits of information). Therefore I can imagine that the attempt to understand the brain's function by observing its interactions is a particularly difficult job.

Ability to communicate is not a yes or no property as is discussed in chapter 10. Foundation of communications is a continuum of behavior that is dynamically segmented by systems in order to find a way how to communicate with each other. This requires an enhancement of capabilities to be expected from communication partners who in my communication channel model are already much closer together in features than in the standard model where information is flowing only in one direction. In the standard model the sender and receiver perform quite diverse activities and implement even quite contrary algorithms like for example encoding vs. decoding. The agent model in use here explains phenomena exceeding the standard model but the differences between senders and receivers are greatly diminished. The design of sender and receiver is basically the same with a small exception: The sender gets its reference for the Δ-function from the outside while the receiver gets its reference from memory. However, the symmetry between the parties can be made perfect by applying a few more changes. In that case a communication chan-

8    http://www.cs.bham.ac.uk/research/projects/cogaff/misc/information-difference.html

nel would not be aware of "senders" and "receivers" anymore. Communication would be the sole result of changes in the error model and its inverse – albeit the names "error model" would be not so appropriate anymore.

# 8 Compression and Error Correction

Most literature includes the step of compressing data these days as shown in figures 7 or 33. This seems necessary because the stream of data to be transmitted is often containing less than one 1 Bit of information for each technical Bit of representation. Unfortunately, the redundant information is not necessarily well "mathematically" entangled and hence difficult to exploit for the purpose of error correction. Since error correcting codes are adding additional overhead (redundancy) to actual informational content, compression is an important and indispensable feature in all practical applications.

Commonly, (lossless) compression is understood as the task to find a code set allowing more efficient use of technical signals. Huffmann-coding, run-length coding, arithmetic coding, Lempel-Ziv coding or Burrow-Wheeler block coding are lossless compression techniques which are accompanied by numerous lossy compression algorithms mainly designed for multimedia applications.

When communication parties engage in unconstrained communications they must not only find out about the spatio-temporal coding, the semantics of the communicated but also get a leg on data masses. In this process selective transmission, data compression and data encryption protocols must co-evolve among them. It is therefore to be expected from a theory used to explain sophisticated communication evolution that it will integrate aspects of data compression.

Even if the ability to compress channel data is not the main focus of this paper I am quite confident that it is possible to treat it as integrated if the communication parties are developing and optimizing contextualized codes with error correction. For this they probably need no additional algorithms than are required for finding the initial (unoptimized) codes in the first place. Prolonged use under temporal constraints should show general preference of shorter, well balanced codes which are signifying compressed streams.

If we agree that the source of strength of error correcting binary codes lies in their structure, i.e. in the implicit relationships among the bits, then there is no good reason to reject the idea that any object transmitted with sufficiently repairable internal structure can be considered as a complex error correcting code.

Particularly complex objects used for transmission, such as large RDF packets, can benefit from compression irrespectively how well compressible is the underlying (let's say XML/UTF-8) stream. How can this be achieved and how useful is this in context of error correction? Several lossless techniques were studied such as stream compression or various ways to compress the graph structure [70]–[72].

Well, one general practice for compression can be the choice of temporary, very narrow context substitute symbols (substitute vocabulary or substitute concepts) which are concentrating on superficial structural features rather than semantic commonality. It seems that this was the motive behind graph pattern compression described in [71]. It is true that announcing such substitute symbols is causing an overhead which must be overcompensated by other structure simplifications before being effective. However, there exist compression methods where the replacement is implicit, such as in the LZW algorithm where the receiver is reconstructing the compression table from the compressed stream. It is therefore not unreasonable to believe that compressed complex objects can be "self-documenting".

Defining such structural ontologies for the graphs can be useful for integrating with error correction mechanisms as error correction relies on predictable structures. This can be exploited under some circumstances: Firstly, the error cannot be explained as a streaming error, i.e. bytes arrive correctly. Secondly, the receiver can actually accept, reject or accept propositions after a modification. This can require a body of knowledge about reconstructed object parts.

# 9 The Integrated View

So far, several key concepts have been discussed in relative isolation. In this chapter I will try to draw a more integrated idea of how a communication channel should be seen, particularly how control applications and communication channels should be seen as projection of a yet more general model.

Communication channels have two very important properties, a capacity and a direction. In truth the source and target points (senders and receivers) also play an important role in many questions such as how they recognize if they are addressed in broadcast situations or how they synchronize with the alleged communication channel.

However, classic communication theory, as was originally proposed by Shannon, is mainly dealing with abstract one-way channels and their theoretical limits, mainly governed by the source coding theorem, the rate distortion theorem and the capacity channel coding theorem. For any more practical application we must chose the way information is encoded (compression, encryption), technically transported (delay!) and last but not least: protected against noise (also encoding). This leads to a more detailed understanding of communication channel as shown in figure 33.

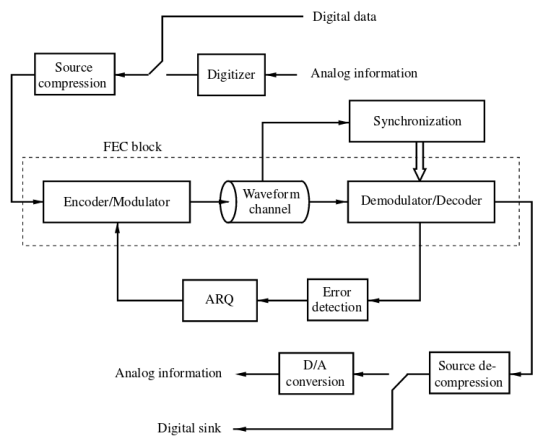

*Figure 33: Two-way channel as described in [62]*

The last point about noise protection is particularly important because all communication channels suffer degradation: Input and output are not the same. Communication without error correction finds only few fields of application these days as even on-chip communications and between-chip communications are using some forms of error correction or at least detection. The higher the frequency of component operation and the stricter the power requirements, the more compelling it is to make some bets on error correction techniques.

Because we are looking for a way to explain how devices can be designed to recognize, exploit, maintain and optimize communication channels, we are trying to understand how error correction can be an integrated part of it because error correction (at various levels of communication nesting) seems to be the key capability in order to map input to some reasonable internal function. According to my opinion, sensible reaction to input is indispensable for the sender to recognize utility and reliability of the channel. This reliable utility will be the condition under which the sender will like to use this channel again.

If spontaneous development of communication capabilities between devices was based on sophisticated error correction mechanisms, ones which were not (completely) specified out as would be the case with solutions resulting from a strict engineering process, the question is then how these underspecified error correction and error detection mechanisms can work?

In order to start developing technologies and devices which can engage in broad range of synchronizing, triggering and regulation activities, we should have something like an high-level view of a communication channel that is semi-bi-directional. I intentionally say semi-bi-directional as we still see decisive sender (source) and receiver (sink) role in this.

I propose to use the schema in the middle of figure 34. It consists of information modulators, encoders, decoders, a source, a sink / target and a noise input to both sides of the channel (sth. that is lacking in figure 33).

For this let us remind ourselves how signal reconstruction can be obtained in two fash-

ions:

1) On one hand it is possible to give up complete independence of signal components[9]. Signal components are then obeying a mathematical entanglement which can be used in order to reconstruct the original signal's structure. However, human language uses no such mathematical redundancy and yet still fulfills the error correcting function. In human languages grammatic, syntactic and semantic redundancy [73] is used to make natural language error correctable. It seems that error correction is not relying on well ordered code sets - Entanglement is the key concept. This concept can be understood quite vividly as a mesh of entities which can be distorted, bended and even destroyed by fraction before the picture which is imprinted on it is irrecoverably lost.

2) On the other hand patches of information can be re-sent in order to fix defects reported by the receiver. In this process a reverse flow is required which allows to communicate the position and size of error. This process can implement a good-enough polling strategy where the sender starts with a transmission that is hugely under-sampled and therefore can never be correctly understood by the receiver (domination of omissions and erasures).

The receiver must identify and prioritize areas of message which must be repaired and communicate it to the sender. The advantage of this procedure is that the sender needs not to have a perfect understanding of receiver's interpretation mechanism. This is an engineering-advantage because usually, the sender must know the receiver's interpretation model in advance and must reason about decodability in advance. This is true for compression, encryption and error correction algorithms alike. Engineers must develop encoders and decoders to the same design sheet so that they are making exactly the same mathematical and algorithmic assumptions – something that we cannot upkeep in devices which must self-integrate.

Particularly for the reverse channel I have drawn striking similarities with closed-loop control (control via feedback): Systems report what they believe what they have received or how they have implemented it and the sender evaluates if he needs to take action in order to correct the implementation. As already discussed, this can be understood as a code set arbitration process.

## 9.1 The Information Modulator

The source is generating a new informational content which is first going into the information modulator. The modulator is not like a modulator known from analog communications (as discussed in other chapters). The modulator here is a component for selecting and prioritizing information to be scheduled for communication given the nature of the change on the input side and given the change of deviations reported on the feedback link. The error detector is a combination of the Δ-function and the error inverse function from chapter 7.

The information modulator is the authority deciding how to fill the virtual message boxes introduced in figure 27. This will result in filling rates over time as were sketched in figure 29. In extreme cases we will observe a behavior of sending distinct messages but any more complex sequence of message bursts and complicated message ping-pong can be realized or explained using the model of senders and receivers running a communication chain between them as was depicted in figure 27.

Information modulators can for example remove high frequency information from stream in order to maintain fluid stream of media while degrading quality. In digital communications these modulators are called adaptive quantizers. Skype users know the effect: When Internet connection is shared among more bandwidth users the picture and sound lose their crispiness. Crispiness slowly returns when bandwidth use can raise again.

## 9.2 Encoding

The content is encoded only after necessary information has been selected for transmis-sion in such a manner that it is compressed and the code is suitable to receiver-side error detection or even correction. Once this is done the box travels with some delay towards the receiver. The longer the time between input and read-out the greater the level of deterioration of the content. Error protection is like a cooling pad that prevents food in a box from becoming uneatable but only for so long, not indefinitely.

---

9 In digital communications the distinct signal component is the technical bit (low, high) which is transmitted but we must keep thinking of non-digital communications as well → *signal components*.

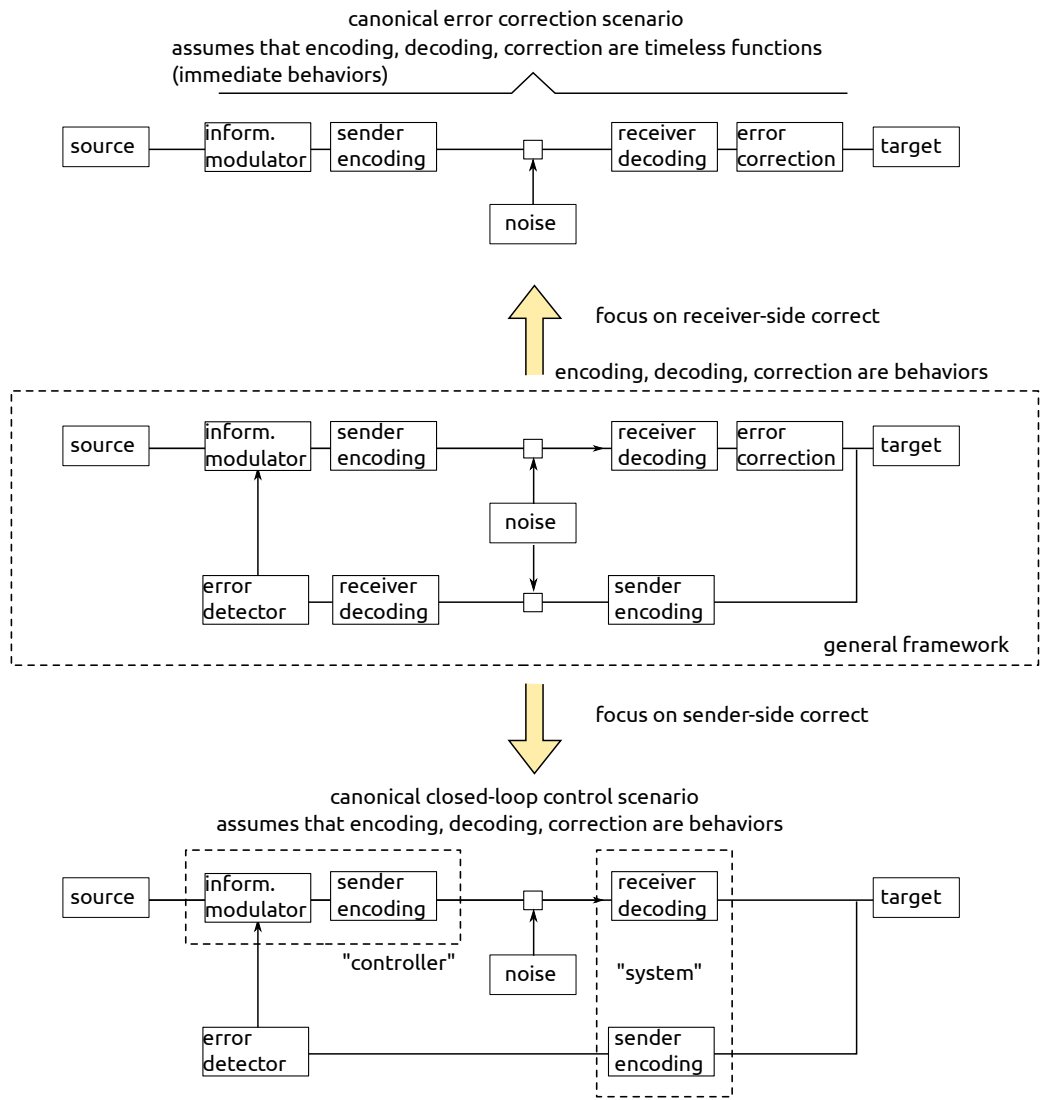

*Figure 34: One-way channels with error correction and control loops are special cases of the generalized two-way error correcting channel (middle).*

Of course other not so well predictable sources of corruptions can interfere, some of them could intelligently plan this (cf. figures 5 and 8). Therefore it is just a nuance in design whether an encoding strategy is designed to prevent random error or crafted error – which can be both credible for the receiver.

Of course, it is not a nuance in design whether a third party is excluded from reading the channel, i.e. its interpretation and error correction mechanisms are facing extremely hostile mathematical adversities when trying to attach to the channel. Systemically, the idea behind read-protection through encryption is to prevent a (harmful) synchronization of a third party (or system) which is not clearly in the collaborative area (cf. figure 5).

In practical applications the question is whether encoding and decoding is a single step or multi-step activity. Where protocols are carefully designed, this is of secondary concern but for devices which have to discover and dynamically, yes maybe contextually adapt their encoding, algorithms capable of performing all of the required which only need to be tuned seem to be easier to implement.

An encoder engine based on neural networks could be just that. Just recently, Google has demonstrated[10] how artificial neural networks are also capable of finding encryption codes. This means that at minimum neural networks have demonstrated that they are capable of finding compressing codes (e.g. [74]), error-correction codes (e.g. [75]) or encryption codes depending on how they are trained[11].

### 9.3    The Transport Path

The forward and backward transport between sender and receiver is mainly characterized by channel capacity (or bandwidth), delay (one element of system's dead-time) and rate of interference (noise).

### 9.4    Receiver's Side

The receiver is performing the decoding. Since the decoded content could be out of range of legal codes (or concepts), the receiver is trying to map the input to a rationally sensible replacement code. If this replacement happens to be the original input then the receiver has realized a perfect error correction.

Since communications are usually nested, a perfect error correction might be not necessary – only a sufficient one which allows the nested communication to correct remaining error and so on – a key idea for lossy compression where the receiver must demonstrate some generative and fault tolerant consumption characteristics..

Since this paper assumes that all physical systems demonstrate temporal behavior as result of particle communications, the cybernetic nature of target could be understood as a cascade of decoding and correcting interactions. Concretely, the TARGET in figure 34 could be not the information entering a robot's controller, for example, but the actual position of the robot's arm – which is indeed the component we want to control.

### 9.5    Handling Feedback Error

If we remain with the robot for a moment, the implementation of the robot's arm position could be imperfect. The receiver reports the current imperfect position and transmits it back to the sender where it is the sender's obligation to decide if the reported result is good enough.

More generally speaking, the receiver is sending a reference back to the sender which is used for identifying the remaining erroneous information. The modulator will schedule and prioritize a corrective update for the receiver.

Needless to say, nested communication will require a nested feedback. It is unclear to me whether the feedback direction can only work like the forward direction where the complete nested feedback is encoded on the carrier and the message on the wire is again a complete message in its own right or whether it is possible send a "fraction of fraction of a fraction" response. I would consider the first solution "standard" and the latter "intriguing to research". Maybe it is indeed possible (under some circumstances) to communicate just an input fragment and let the sender itself find out if that fragment is indeed a fragment intended for correction by a nested error detector of the sending side.

### 9.6    Projection 1: The Single-Edged Channels

Particularly in systems where the target is unknown and its identity is going to be established only after the message has been emitted, it is very difficult to have a feedback channel. Its existence basically relies on the possibility to define a dedicated channel between the two communication nodes. That's the reason why radio and TV have no feedback channels – they are broadcasting and senders do not discern between individual realized "connections" between receivers and the sender. For such cases the classic single-edged channel is a suitable model.

---

10  https://www.newscientist.com/article/2110522
    -googles-neural-networks-invent-their-own-
    encryption/

11  I agree that there is not THE neural network and that they vary be design and learning algorithms. However, ANN always hold the promise to be mixed into more general designs.

In regard to the general model only the reverse path is removed. The role of the information modulator is reduced to a static one where it acts as a sending filter. It could be used like a firewall preventing emission of certain types of information.

Since single-edged applications need not be so much care about delay but are usually very concerned with the amount of noise and its ratio to the net signal, delay times are not modeled for the transmission lines. That's because it is often irrelevant whether a TV show will be delayed by 1 or 2 seconds but it will be significantly impaired if the picture is blurred, noisy, holds digital artifacts or if the sound comes interrupted (worst case).

Single-edged real-time applications are very much concerned with variable channel capacity which can go as much down as zero. Depending on the variability model of the channel's capacity (or bandwidth) receivers must plan in buffering and this buffering is adding to the regular signal delay. In figure 27 this buffer is indicated as at least one input buffer unit. At minimum this will be a single bit of data in digital communications or a single sample from an A/D converter in analog settings. Of course, analog circuits can store analog values (analog computing) and will yet still obey the described schema. Despite that buffers are lacking in figure 34, they are implied to exist – please take note.

### 9.7 Projection 2: The Control Loop

In control application the temporal characteristics play a key role. Here, delay and speed at which the receiver can implement the message influence the strategy or behavioral policy (or dynamical transfer function) of controller. Control applications are very often analog but real implementations rely more and more on digital computers which have a limited temporal resolution (computing cycle) and limited signal resolution. More complex control systems will exchange complex signals, e.g. a sentence like "do a little bit more than you usually do". Despite that the receiver did not receive a numerical value and that the sentence is highly ambiguous to outsiders, such sentences are commonplace between intelligent agents who can convert it into an executable instruction.

With this said it should be clear that control applications can be based on codes and messages which are not immediately associated with quantities processed by the system (constituted of sender, receiver and maybe other parties). It is engineering pragmatics that control models are explicitly processing codes from the real numbers set. These pragmatics allow algorithm developers to define rigid methods for controller design and stability conditions.

In control systems it is believed that the target system is the source of the noise (distortion of signal, etc.). Whether the noise is result of communications or the target system is just a matter of how you draw the boxes in the model of figure 34. Control designers would model a noisy communication channel as an explicit system between controller and the target system which is suffering from its own type of noise. Fair enough.

However, error correction is deemed absolutely the sole responsibility of the controller which is why the feedback channel is critical to most controller designs[12]. For this reason an error correcting element is missing on the receiver's side. Practical applications are not concerned too much with this restriction: They are often willing to implement an internal error correction loop (nested control) in the receiver in order to simplify the design and operation of the outer controller.

---

12 There are of course open-loop controllers without feedback loop and they will care about delay on the line in contrast to many single-edged applications.

# 10 Establishing Communications

## 10.1 Why communicating?

If devices are operating in an unconstrained environment then first challenge is to detect objects with various degrees of autonomy [76] and hence various degrees of nested control [77], the second challenge is to find out what they can do and the third challenge is to invoke these object's functions reliably. A naive interaction with systems with nested control will usually result in no effect, diminished effect, or can, if the other objects model your device's behavior, even become hostile to your device's agenda, i.e. choose strategies which actively suppress success of your activities (cf. figure 5).

Hence, the purpose of communication is to overcome lower levels of control and to address only the right level of control where a reliable effect can be achieved, i.e. to overcome eventual self-correcting and self-protective behaviors of that object. Sufficiently sophisticated control systems can find co-ordinating strategies which allow them to pursue their own agenda despite goal incongruence and at the expense of having to communicate (which is quite some expense). Example of this behavior would be hand waving between drivers who want to pass a narrow bridge in opposite directions.

Addressing the right level of control is functionally decisive. For example, it is sufficient to push a stone in order to move it to a new place as it does not have another nested level of control. This is a sufficient "communication" with the stone even if there is danger that the new place could evoke a dangerous rolling dynamics. However, pushing a robot with position control to a new place will show no durable effect as the robot will return to the old place as soon as the exerted force is reduced. Once the robot is told (by communicating with it) to move to a new coordinate, there will be even no need to exert any force at all.

Of course, the necessary communication could require communication at various levels. Imagine that the robot is stuck in a pit and cannot get out by itself. In that case exerting force in the right moment when the robot tries to reach a place outside the pit would be an example of a synergetic multi-level communication between systems.

## 10.2 Spectrum of Behaviors

The main question is how devices can recognize which behavior is communicative and which not – from the perspective of any involved party. It seems that the key to understanding communication is the discrimination between functional and non-functional behaviors. In figure 35 we see a pair of hands kneading bread. The average observer would not interpret this as a behavior which is intended to act communicatively. Its actions are functionally intended.

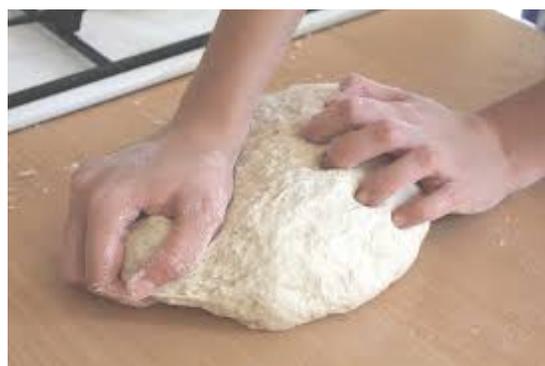

*Figure 35: Functional behavior not suitable for communication (e.g. hand states result from specific activities such a making bread)*

In figure 36 we see hand signs. These are also behaviors of hands but we clearly identify them as signals with the intention of communication.

How can that be? Despite that both behaviors and hand movements can have rare contextually inverse interpretation as communicative or dysfunctional, it is sufficient for two systems two detect that in general a certain behavior is functional or dysfunctional. In order to make that discrimination the object in question must model more functionality of the remote object than is necessary to engage in communication. That is exactly the reason why engineered objects do not have this ability.

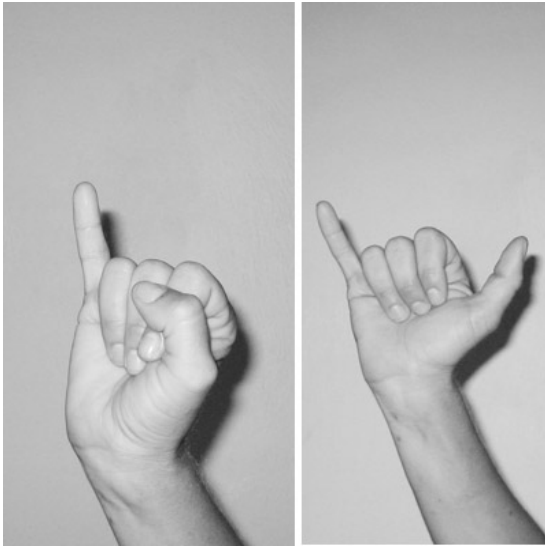

*Figure 36: Dysfunctional behaviors suitable for communication (e.g. hand poses and movements can represent internal states or concepts)*

In figure 37 a segmented spectrum of observable behavior is depicted. The segmentation in green (functional) and red (dysfunctional) represents the observer's believe about the observed object. At minimum this segmentation is representing green (known) and red (unusual) behaviors of which classification object might know nothing about.

As the observed object is performing more and more activities which the observer can relate to any particular function known by the observer at its outmost level of control, the observations are added to the green spectrum of behavior. Otherwise they remain in the red area.

This dichotomy is an intentional oversimplification as the number and granularity of controls is arbitrary. There is no guarantee that functionality found in one level of the observed system is assigned to the same level in the observer system. That means, that there could be more colors used to segment behaviors and that the term "functional" and "dysfunctional" are relative to a particular level of control in the observer.

It is important to note that all behaviors are generated by the outer shell of the system (body and body controls) but some will have a greater proximity to higher level control's state as is shown with the red circle (nested control) inside the green circle (body control) in figure 37.

It is clear that it is dangerous to model the green and the red behaviors only on the grounds of frequency because once a system can and must communicate, its behavior could be predominantly communicative (red). At first, when systems did not yet develop a communication relationship, modeling functional behavior based on observed frequency is just and helpful and special behaviors could be based on anomaly detection.

## 10.3   Rigid Fixed-Protocol Objects

For reasons of efficiency and speed, artificial objects come with pre-built protocols and communication channels which can be exploited by compatible devices without any modeling. This requires high levels of compatibility at all levels of control and protocols.

Incompatibilities can cause partial or complete lack of function or dangerous behaviors in certain situations. For now, our goal should be merely to explain how two systems / obects / devices could establish and maintain communication at all.

Many experiments in the area of configurable robotics are relying on a predefined communication protocol between the robots such as blinking lights, sounds, WiFi signals, pictograms, screens or peculiar movements. Based on these cues the objects implement a defined protocol with defined expected functions and defined triggers to invoke them (a "configuration"). This allows researchers to study the utility and robustness of such configurations. However, it is not known whether these configurations are the best ones possible. Self-integrating systems hold the promise to discover these communication configurations faster than researchers can and to remain flexible henceafter, i.e. evolve communication given changing conditions.

Flexibility should not be underestimated. For example, engineered objects (the robots) usually come with an a priori knowledge of which cue is green (functional) and which one is red (communicative). For example, designers of robots who design flashy light patterns on a robot's body for communication will make sure that special flashing behaviors are recognized as communicative by its peers (all of same species). Therefore the spectrum in question is a species-specific property of those robot objects and those robot groups can work perfectly until the day when they meet another species of robots who use flashy body lights for quite other purposes. This can end in a total failure of function for both robot groups.

## 10.4   Intelligent Observers and Rigid Objects

The flexibility of the situations could be improved if at least one object possessed extended communication flexibility. Such an observer object can learn to establish a colored subsegment of the red range and to assign it to a particular level of its control.

For the flexible observer, the red area is defining a set of behaviors with a signaling property. If the observed object is artificial and does not model external object's behaviors, it could still rely on that ability for communication. For example, consider the robot in figure 37: Its designers could define virtually useless body configurations which will not occur in usual working conditions. For example if the robot folds up vertically, it could signify a lack of input signal after a tolerance time. It could also report errors by starting very small (safe) swinging movements in order to indicate that it is confused or simply to say "No." A human observer would immediately recognize these behaviors as communicative as, according to his best knowledge of the robot's function, they serve no immediate function.

In some sense we can say that the red spectrum of behavior is a reserved code which can be used by the nested control to directly express its state. If the observer acclaims this code as a known symbol then this will establish a connection between the nested control of the robot and the nested control of its observer. This saves the observer the time and effort to interpret undesired but possible normal behavior and to hypothesize about its reasons.

In figure 38 we see a conceptual example what this could mean. Let us assume that the observed object is performing a cyclic function which is observed and modeled. After a certain amount of confirmation has arrived, observing more states from the functional spectrum will not yield any insight regarding system's further behavior. A change in direction would be completely unpredictable. Only if the system is suddenly moving to one of the two external green positions shortly before direction of movement reverses then it would have a communicative role for the observer. For example, the observer could reset an internal timer and reverse its function in exactly the same moment as the observed object.

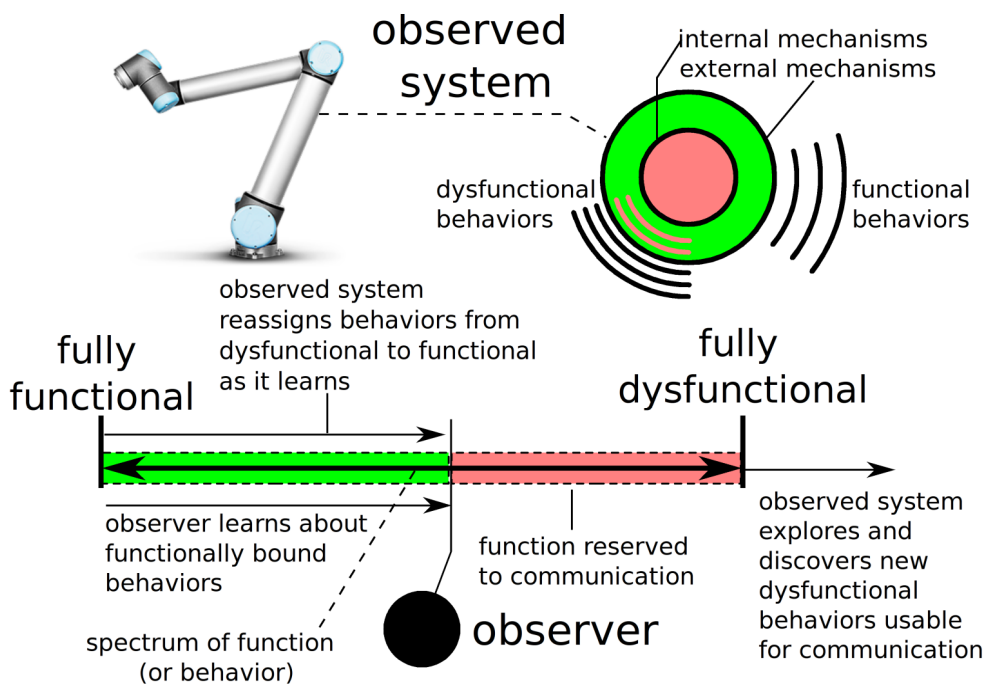

*Figure 37: A learning observer models functional behaviors of observed objects and filters them out of the range of possible communicative behaviors. If the two objects can agree to use dysfunctional behaviors for invoking nested control functions then they will engage in communications.*

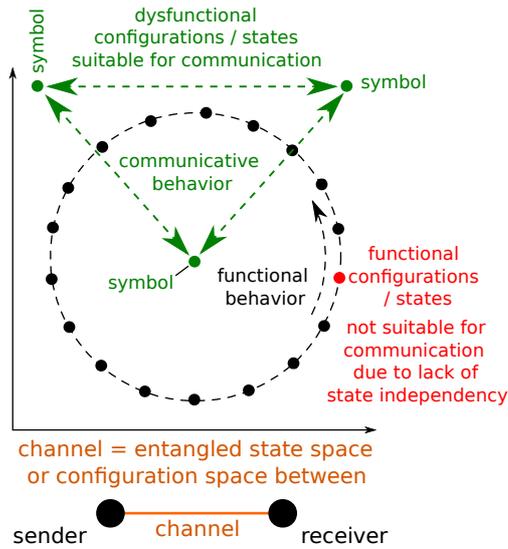

*Figure 38: The coloring is inverted now in order to indicate that a cue is suitable for communication (green) or not (red).*

So far the communication has not included intervention by the observer. Imagine that the observed system accidentally falls into the center and does not move from there. Let us further imagine that the observer has now the ability to interfere with the state of the observed object. For example it could deliberately push it into the center in order to turn it off or it could move it into one of the corners in order to initiate a change in direction. In that case we would say that the observed object is exposing an interface for the observer.

## 10.5    What is Input and Output?

Clearly, the above example is heavily relying on a state-space concept of the system which I believe is applicable to almost any system as I have explained in [10]. The direction of communication in this example relies on the ability to exert influence on the system in such a way that the nested control is taking notice of it. Communication interfaces are the easier to access the less force needs to be exerted on the system. Ideally there are additional dimensions making up the configuration of the observed object which have little internal inertia and hence are more suitable for receiving input from the outside. So, a system can be said to have an input interface if it offers a configuration dimension of which it is constituted which exhibit little inertia and ideally are not involved in functional behaviors. This would be the case with a light sensor: It requires a relatively little amount of radiation in order to change its state and because it cannot emit any influence it is functionally irrelevant. Hence it is an ideal candidate to serve as an INPUT to a higher level (more deeply nested) control.

Theoretically, systems with controllable stiffness of their configuration can control which of their parts exhibit much or little inertia. This allows them to open and close an interface depending on which object is trying to use it and in which context.

In contrast to a low-inertia interface we can have infinite inertia dimensions of the object's configuration. Such dimensions are ideal for sending signals as they are immune to noise. More generally, such dimensions can act as controllers to the object's environment. An LED is example of such a dimension. Its activity is only depending on the nested controller's internal state and no (reasonable) amount of interaction can make it go off or go on.

Please note that low inertia dimensions (sinks/targets) and infinite inertia dimensions (sources) are best suited for communication if they play little role in the object's functional activities. However, this input / output thinking is just two extrema in a spectrum of interaction realized through system's configurational dimensions as shown in figure 39.

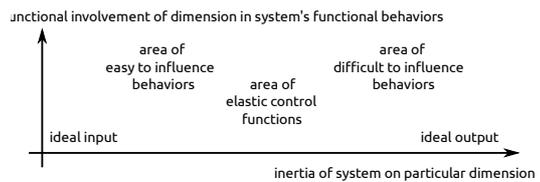

*Figure 39: Spectrum of interfaces depending on inertia of dimension and functional involvement*

## 10.6 Development of Communication in Intelligent Object Groups

The most interesting case is when both objects are intelligent and can actively contribute towards defining a communication channel and usable encodings. Figure 40 shows how the communication channel would be established. It follows the basic steps first indicated in figure 6.

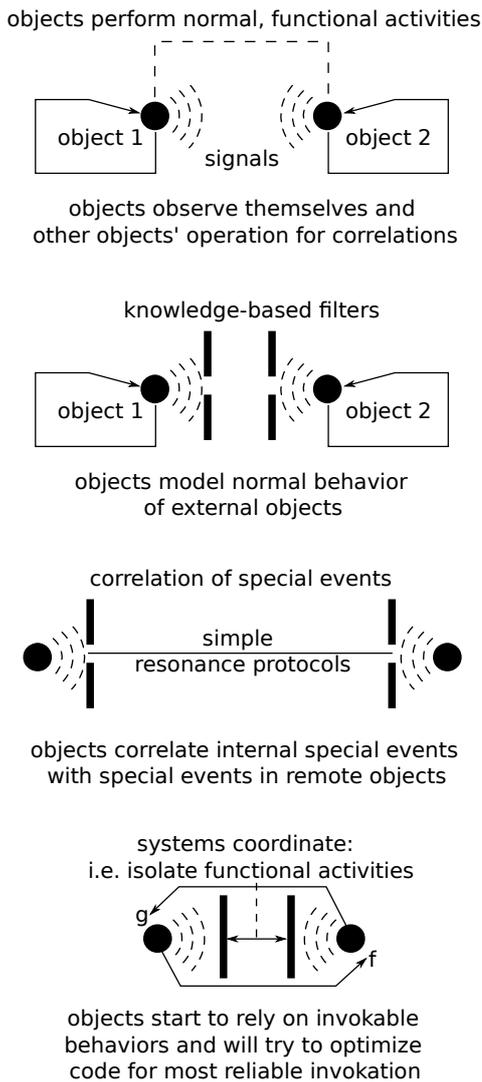

*Figure 40: Schematic evolution of communication between intelligent objects*

At the beginning of this sequence, objects are unaware of each other and only influence (but do not communicate with) each other. At the end of this process, objects have eliminated systematic error from their interactions.

Step 1. Two objects are agnostic of each other. They perform their normal functions. The objects observe their own performance and correlate any deviations of operation from their expectation with events observed in their environment. This process can bring forward attention to other objects.

Step 2. The two objects want to channel the effects observed in relation to the other object. For this purpose they create a mask (possibly from episodic memory) which is used as an attentional filter to detect just the right cues necessary to improve prediction of own performance.

Step 3. Since objects observe and model all potential causal sequences they detect that they are entangled in some way with the other object. If own activities cause desired reaction on the remote object then they are memorized with a trigger. If objects are capable to incorporate new triggers and new behavior sequences into their policies then they can start making the impression that their relationship is characterized by a communication relationship via simple protocols based on short sequences of calls and requests.

Step 4. As part of the object's attempt to gain and retain autonomy, the objects will start to actively inhibit all low utility channels between them and to iteratively reallocate useful communicative behaviors out of space of functional behaviors (where possible). If there is no resonance attained in this way then the process of isolation and suppression will continue to the point where either the two objects never interact or to the point where one object inhibits action of another object to such a degree where it cannot act. If this suppressed object's existence conditions underlie some real-time requirements then this condition is "deadly" as it will violate those conditions and will disintegrate. Such is the source and nature of systemic aggression which is reversible as long as the model in the suppressing object's controller can accommodate new utility for the suppressed object. Rogue objects are defective and attempt to inhibit other object's potential to realize its function irrespectively of the potential to usefully communicate with it and benefit from its function. The more ontological modeling is involved in this setup the more likely it becomes that certain objects are generally considered "bad" or "good" and hence become either generally suppressed or generally communicated with.

## 10.7 Establishing Robust Communications

Once two intelligent objects have detected which of their properties are entangled and which useful patterns (codes) are observable with them, objects start to optimize their behaviors in such a way that "natural" behaviors used for function and "artificial" behaviors used for communication are maximally dissimilar from each other given some base (frequency, matrices of pixels, wavelet coefficients, statistical measures, etc.). This is similarly motivated to the random sampling approach in compressed sensing [78].

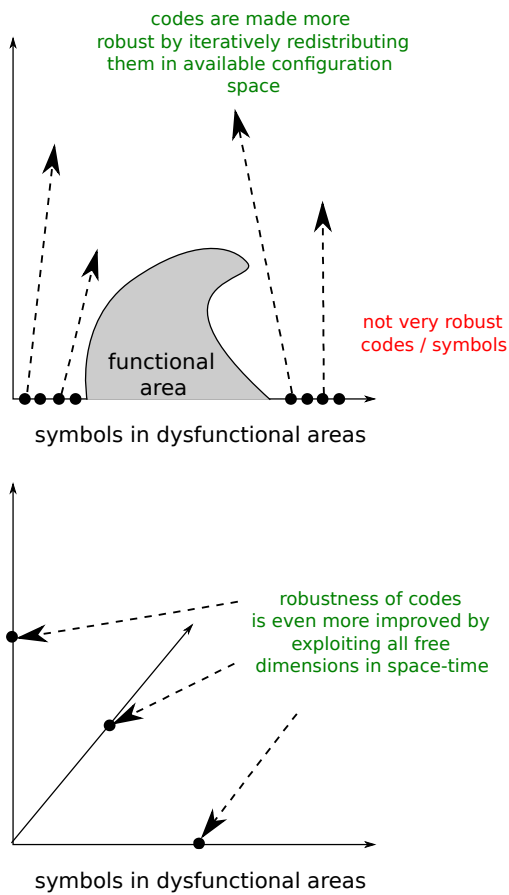

*Figure 41: Process of improving reliability of communication leading up to long temporal codes (as time is the only freely extensible source of additional code dimensions).*

Figure 41 is showing how the objects' models could evolve after the initial phase. This figure shows a very high dimensional space which number of dimensions is a product of time samples and degrees of freedom in object's configuration.

It is to be assumed that functional behaviors of the objects will cover an unknown subspace, the "functional area", with the total space. Further we can assume that the object has found outliers outside of the "functional area" which have been found useful cues to predicting evolution of states occurring within the "functional area". Such useful cues are codes in terms of communication.

It is reasonable to assume that such signaling prototypes are at first nearby the "functional area" and since the modeling in high dimensional spaces is a challenge in its own right, it is to be expected that functional behaviors will eventually "overspeak" the cues leading to reduced reliability of them. In order to reduce noise (and hence error) on these cues, they must be moved further away from the functional space. Once the process has been ideally accomplished, code prototypes are dwelling on dimensions which are completely unrelated to observation of action. We would call them highly artificial where other behaviors would be naturalistic.

The number of codes necessary to coordinate or exploit all the functions in the interacted with systems can be higher than the number of physically available free dimensions of behavior. If there are remaining free dimensions then those can be replicated in time in order to emulate a higher dimensional space for communication prototypes (which serve as cues or codes). As consequence, communication must obey certain sequencing rules in order to be effective: Receivers and senders must have a convention about which dimensions are presented in which order.

Clearly, the reallocation of cue prototypes is only possible if both systems are modeling possible cues as codes. Since we assume that systems can directly interact using the entangled object properties, this can be exploited to push the cue in the desired direction by each side as I will discuss in next chapter.

# 11 Layered (Nested) Communications

## 11.1 What is nesting?

In chapter 9 we have seen how a double-edged communication channel could look like and how it relates to single-edged channels and feedback control. This was a "2D" perspective.

By including nested communications into this model, we will get a "3D" perspective of a communication channel (cf. figure 42). In chapter 10, a basic nesting of control was used to explain the nature of communication and its evolution. This nesting poses higher challenges to communications as more interactions have to work together and there are more reasons to fail [41]. The model of the eight incompatibility types from [41] will be later used in order to assess the potential of comprehensive error correction approaches to overcome them.

The model in figure 42 is a little simplistic as the majority of applications requiring layered communications is doing this in order to exploit a single simple but physically extended mechanism (the "communication channel").

Figure 43 shows an extended version of the concept in figure 42 where each layer is holding two nested mechanisms requiring communication. Please note that organization of senders and receivers is the same.

The question is whether error correction is extended by looking at nested mechanisms and nested communications? In order to explore answers to this question I have expanded the concept from figure 27 into a full fledged bi-directional example with three layers of communication as shown in figure 44.

The most concrete communication layer in this figure is colored red. The first nesting is yellow and the nesting into yellow is colored green. This example is not holding competing communication parties but the reader should imagine parallel branches in places where mux and demux blocks are placed. These places represent a switching layer in the device's overall architecture.

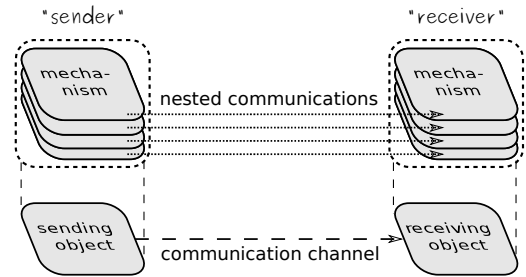

*Figure 42: Nested mechanisms in objects result in nested communications. Communication is only possible if there is a receiver at the right level of nesting interpreting the signals.*

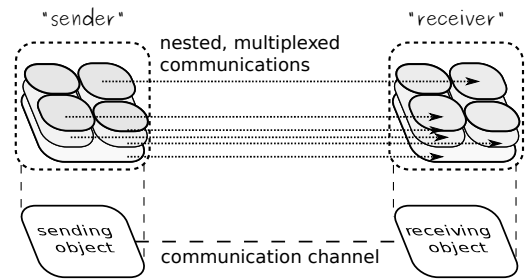

*Figure 43: Nesting involves multiplexing of signals (routing of communication) in almost all practical applications.*

In this model it is assumed that classic forward directed error correction (e.g. use of error correcting codes) result in a weaker error function on receiver's side – such forward error correction is not explicitly modeled in the example shown in figure 44. However, this explanation does not entirely work for the reverse direction and hence I admit to abbreviate the matters somewhat.

Please note that the full-duplex configuration of the example in figure 44 is showing four edges of communication between the two blocks representing two endpoints of communications. In real technical applications this would result in two stream directions and each stream would hold the regular forward edge and the corrective backward edge of the other communication direction:

Stream 1: LR-forward + RL-backward

Stream 2: RL-forward + LR-backward

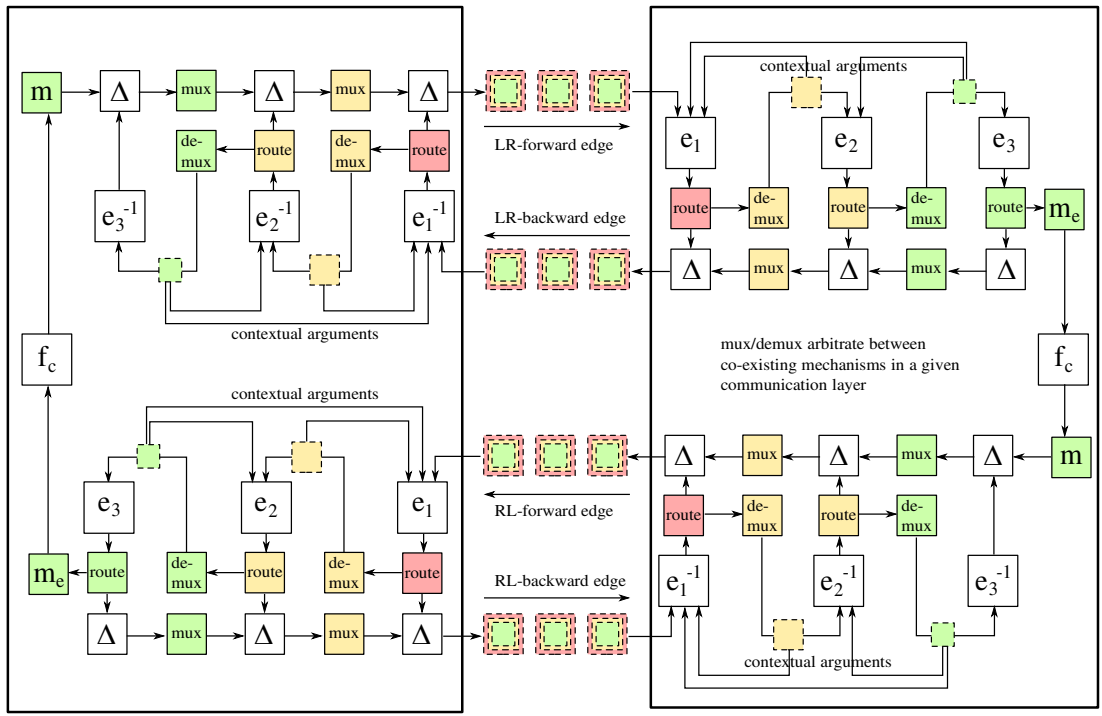

*Figure 44:A bi-directional double-edged channel (four edges) used by three layers of communication. In technical systems of today, the two communications edges (forward/backward) are mapped onto one technical path of communication yielding two paths (full duplex) where each direction carries forward and backward edges. This model requires error correction for correcting routing. Packets can influence corrective behavior by feedback links and by this they can be designed to actively support correction of routing errors. However, other more direct means should work better. The function $f_c$ is a core function which is performing the system's main objective. In autonomous systems $f_c$ is the core autonomy function.*

Nesting of communication is requiring that error correction is performed in sequence. The first communication layer is giving the first try to correct error ($e_x^{-1}$) and then to forward the content to the next error correcting function. As already suggested, error correction might not be perfectly effective. In that case the purpose of the router is to decide how good the content is after correction and not to pass it on until better quality has been accumulated on the red layer. In trivial cases it is simply an unpacking station where yellow virtual boxes are unpacked from red virtual boxes or where green virtual boxes are unpacked from yellow virtual boxes (splitter).

The multiplexers and demultiplexers always require two inputs: the signal and a selector. The blocks shown in figure 44 are only having one input. This implies that selectors and actual signal content must be transmitted using the same means. In technical application a "header" is usually attached to the "payload" in order to provide the "selecting" information for the demultiplexers. On the sending side the mux-blocks are doing the opposite: They add the "headers" to the "payload" in order to signify where it came from. Clearly, the content of the "headers" has a symbolic nature – an address. In cases where there is no "payload" or very little of it, information transmitted between objects (the nested virtual boxes in the middle) can be seen as a complex package of addresses. Colloquially, we would speak of a *symbol* with complex connotations and semantics rather than addresses but the true nature of a symbol is its addressing property in communications.

Since addressing information is transmitted along with the actual payload[13] it is susceptible to corruption as much as the payload. Therefore it is necessary to have means to

---

13 which can be theoretically decomposed into address information completely, i.e. payload consists of addresses only handle mistakes in addressee selection. A false selection will result in a radical decrease in quality of received information. This would manifest itself in the inability to correct errors in large portions of the received information. The minimum response to a radical degradation of information quality is to choose other error correcting parameters for the previous error correction blocks by providing additional contextual arguments. Application of contextual arguments is an easy concept only if no multiplexing is applied as in fig. 11. Otherwise the contextual arguments compete with each other and the resolving error correcting inverse function must implement some means of arbitration.

## 11.2  Temporal Codes and Protocols

As a matter of general experience, outer communication layers usually use simpler codes than the inner layers. In digital communications the outermost layer is transmitting a symbol from the set {0,1}. Even if we assume some parallelism of such signal lines, this means that all more complex entities must be expressed as sequences of such elementary representations.

In figure 44 the only place where the completion of a sequence pattern can be identified is in the inverse error model block which has an input buffer to buffer a sequence of arriving codes. In order to recognize particular temporal sequences, many implementations in software rely on finite acceptor state machines or *lexers*. Such algorithms are usually not fault tolerant as they will accept only the one correct grammar and they will do so only after a complete sequence was recognized.

Here again we have the word *recognized*. Recognizing fuzzy sequences, i.e. sequences with high potential for syntactic error, requires more tolerant mechanisms than algorithms based on automata theory. There are several more fault tolerant methods in the pattern recognition domain. For example, Partially Observable Markov Decision Process models (POMDPs) have a good record of success in auditory sequence classification.

## 11.3  Allocation of Error Correction to Protocol Layers

Before we can speak of allocating error correction mechanisms to nested communication channels I want to quickly discuss the idea that any two communication partners usually use more than one layer of communication. That is they do not interpret the raw input signals as the primary data but use several steps of interpretation in order to retrieve the actual information. If this interpretation is trivial as is in network protocols (just headers are decoded in order to manage data forwarding and assembly) it is easy to oversee that any nested protocols are the equivalent of nested communications. For example, a communication between two IP-parties is a nested, virtual communication between Ethernet nodes which have real communications between places identified by MAC addresses. Another example is the data received by the web server who is not consuming the data itself but forwards it to a complex, nested architecture of service providers and data consumers.

The allocation of corrective mechanisms in the layout of communication stacks could be suboptimal if error correction is falsely understood as sole responsibility of the lowest digital communication channel. I want to emphasize the various options to lay out error correction in communication stacks. In order to raise attention to allocation options I will rely on the idea that communication channels can embed communication channels like Matryoshka dolls.

Usually, consideration of data communications ends at the application layer in the seven layer OSI model. However, this particular model is not necessary the full truth in communications. In fact many protocol stacks do not offer exactly one protocol per OSI layer and completely different models with layers of other responsibility are imaginable. Moreover, complex middleware based on sophisticated component models and sophisticated application architectures based on them can add any number of additional layers of communication beyond the usual seven. There is an unbroken trend to add communication layers in order to improve manageability of tech-nology in different business domains[14]. Different professional groups have different (sometimes contradicting) requirements towards technology and the solution to satisfying the various needs is to separate their concerns by layering communicating systems which may obey different paradigms of administration.

Of course, this increases the difficulty to achieve compatibility as it will require that either a perfect match exists between the layers on senders' and receivers' side or that sufficient convertibility of underlying communication protocols can be achieved. The latter is the more difficult the stronger a certain stack relies on certain responsibilities of a certain layer. Changing the protocols may fatally disrupt correct operation of higher level protocols. For example, if a protocol with session management must be replaced with a different protocol that has no session management then the responsibility for managing sessions has to be taken over by a higher level protocol. Such reallocations are usually very difficult. One way to get around the problem is to introduce several layers which can redundantly manage communication sessions. This would at least give an opportunity to build a gateway that translates between sessions at different layers.

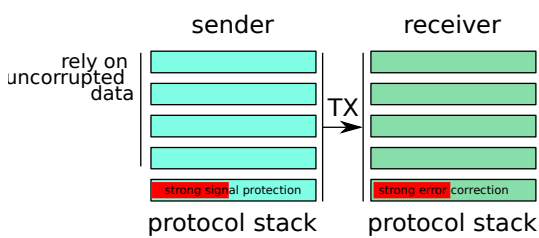

*Figure 45: Allocation of error correction to the digital data carrier protocol.*

Similar allocation questions are related to error correction. In digital communications error correction and error detection are among the first chores which are assigned to "digit transferring" protocols (cf. figure 45). Such protocols can be physical (e.g. use extra bandwidth to send corrective data) or computational (e.g. use extra bits to send corrective data). Please note that for sake of generality all following figures have no concrete technology labels attached to them and that the number of layers is considered arbitrary.

Usually, technical communication layers are designed to be very efficient. The amount of overhead produced in comparison to transported net data is relatively small (even if there are exceptions). However, growing the number of layers and the use of verbose protocols can shift this ratio into arbitrarily inefficient proportions.

For example, imagine that a task requires transmitting a binary value ("yes" and "no") but this information is highly encapsulated in a complex RDF/XML structure. The overhead consists of several KiB of data: The headers of the network protocols, the BOM of the UTF-8, maybe several irrelevant extended length characters were transmitted, the XML should contain namespace declarations and could contain several long and nested tags necessary to encode a minimum structure to hold the transmitted message. Last but not least, the UTF-8 encoding of "yes" and "no" is holding a lot of redundant information which can be used to restore the information: A "ye", "YES", "es", "yess" or "ja" could be mapped to 1 and the same for the *no*. In special cases detecting fragments of data with above properties could be enough in order to decode the message even if other formal parts could not be read. Implementing such a receiver would turn the error correction model from figure 45 upside down as shown in figure 46 while making it a lot more efficient.

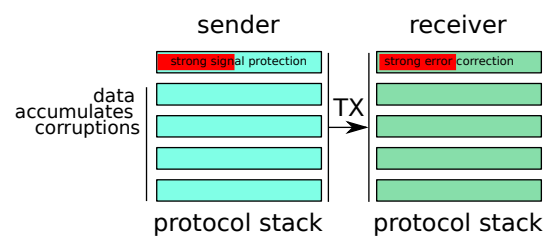

*Figure 46: Extreme approaches to error correction could accumulate all data corruptions until fixed by the inner-most interpreter.*

Transmitting data according to the model in figure 46 requires strong motivation, as sophisticated error correcting mechanisms must be laid out for this purpose. Possible explanations could be (like described above) extreme

14  http://www.ripose.com.au/ExtendedOsiModel.html

mis-proportion between net and gross data needed to transmit or limited (or absent) error correction mechanisms on fundamental protocols. However, more balanced distributions of error correction can be conceived. Figure 47 shows a perfectly balanced responsibility for correcting errors. This is similar but not the same as convoluted codes because in the layered design information is indeed processed between applying error correcting mechanisms[15]. Such models can either capture specific types of errors or they can overlap in responsibility.

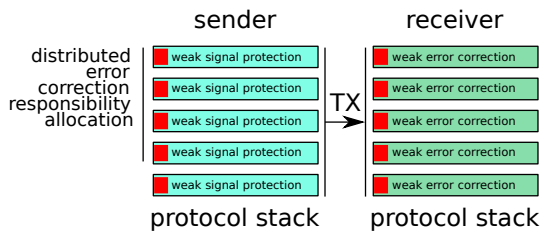

*Figure 47: Even allocation of error correcting capability among nested protocols.*

In case 1, for example, the bottom layer would correct bit flippers, the middle layer would correct incorrect XML tag positions while the topmost layer would correct illegal attribute assignments which are not allowed by the respective business logic.

In case 2, all layers would assume to take into consideration some unfixed errors from the protocol layers below. The topmost protocol would implement error correcting mechanisms which are taking into account that some characters could be badly coded, that some XML formalisms are violated or that values do not perfectly match business logic constraints.

Ideally, error correction is eliminating any observed error the first time it is observed in any of the protocol layers. Most digital protocols are designed this way at the expense that if error rates are raised above a certain level, the error correction failures will culminate in a complete denial of service. RX errors are reported to sender (if possible) , the receiver administrator and the processing stops. Ultimately, even a small fraction of communica-tion errors could lead to a complete blocking of channel if the error correction proves too weak.

In pursuit for better grace in degradation protocol designers could want to accept certain amount of remaining error to be passed on to the next protocol layer. Passing on such errors is not without a hitch: Even a small error in a non fault-tolerant embedded protocol can lead to a complete misinterpretation of content. This means that the factual result of error correction can introduce a new error in the nested channels. Worse, a seemingly greater error at a lower protocol could be theoretically better understood and corrected than the introduced error in the embedded channel. Therefore, combining protocols with error correction capabilities and tolerant error correcting behaviors requires careful layout and parametrization of stack, similarly to filter design (cf. fig. 48).

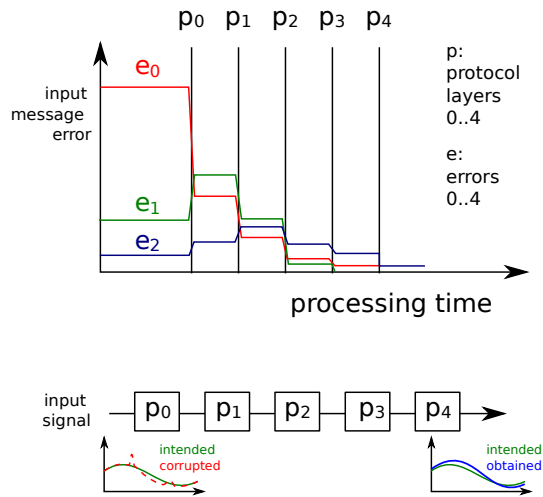

*Figure 48: Error correction can introduce errors in nested protocols (or derived signals) which must be corrected in the next layer (next filter). The outcome of this process could leave a remaining error for the receiver to process.*

Beyond that, individual stages could need to communicate with each other in loops in order to minimize the total amount of error as is true with for example Turbo Codes [14]. Turbo Code decoders have a feedback loop between second and first decoder in order to "tune" the first decoder's function yielding better decodable input to the second decoder. As I have already discussed, this can be seen as a contextualized selection of code sets.

---

15 Error code convolution is a special case of layering error-correcting communications where no other processing of the data occurs.

# 12 Towards Implementations

It should have become clear by now that creating self-integrating environments will demand intelligent features of the devices. My basic appeal is that objects must be designed in such a way that they can communicate depending on the functional needs of the integrated function. If properly designed, devices can flexibly exhibit signaling, streaming and regulation in their communications.

Furthermore, it seems that pattern recognition approaches are particularly well suited for implementing self-evolving codes and should be a central cornerstone to proper device design. The question is then in which way these codes can evolve into error correcting codes. Well, code detection based on pattern recognition is always corrective in general sense (or tolerant) and error correcting codes are simply well balanced pattern prototypes which are usually spaced using regular concepts (e.g. modulo sets of integers).

It will be the nature of spacing, context and purpose of use whether such codes can be considered error correcting, encrypting or compressing – or any mixture thereof – but it will be the flexibility of system's architecture which will define in how far this theoretical possibility is technically exploited. The following subsections are concerned with the question how devices should be architected in order to support as much self-integration in an environment as could be feasible for practical IoT-devices and other intelligent devices produced by or for the *fabrication web [79]*.

## 12.1 The All-Channel Approach

Any device of concern in this work has a proper system boundary or "object surface" which can be influenced and perceived by other objects and which is accommodating actors and sensors. As explained in section 10.5, passive hull, actors and sensors are extreme points in a channel type continuum as shown in figure 39. For a technical designer it is easier to operate on these extreme ends and to know which device components will have which basic effect. An actor (or sender) will emit energy and influence other components, a sensor (a receiver) will receive energy and will be totally influenced by external components and passive objects (the hull) will neither sense or receive energy or influence information.

The lack of intermediate communication modes makes it difficult to design elastic, evolving communications as the translation mechanism in such devices must perform a lot of abstract, representative operations in order to produce a valid reaction. This computational overhead is the direct result of the prevalent extremely single-edged communication modes. Devices processing signals at high speeds or with tight energy consumption envelopes will employ analog solutions and, if problem solving is involved, on resonance approaches. However, these are special technical feature [80], [81] which are usually not found in commodity items.

For our purposes, any sending or receiving facility is forming the external hull of an object. Communication channels are attached to this *system hull*. Ideally, a system is a device which can act as an arbitrary router between any of its four-edged communication channels (4e-channels as in fig. 49).

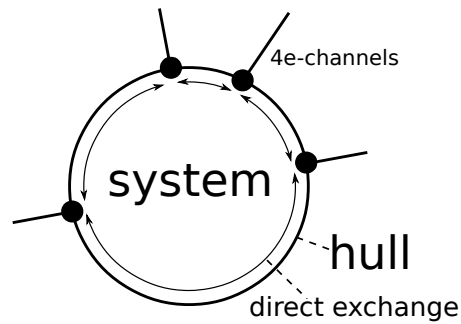

*Figure 49:A system (a device) has a conceptual hull on which 4e-channels are "attached".*

For any practical device, this means that we ask for the ability to receive data on any of the input links (camera, buttons, touch screens, microphones, etc.) and use any display, LED or network device as an output link. If the hull can become part of communication (e.g. use a channel-free side to <u>inhibit</u> input) then it can be modeled as a channel with very high noise, very strong signal damping or very "stiff" line of interaction.

Inside the system we find functionality which can play a potentially relevant role in joint integrated functionality of the environment. At minimum, this functionality is to repeat data from one channel to another. This can require a mapping between codes used by each of the channels. This gateway might work imperfectly, that is, it cannot convert codes from one channel to another without some semantic loss.

Inside the system there must be a switching layer, as shown in figure 50, which is capable of deciding which channel will reach which internal function and which internal function is able to send information.

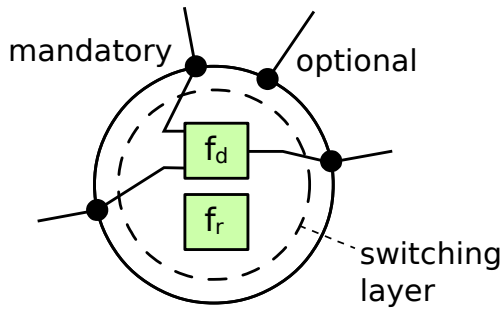

*Figure 50:An intelligent IoT-device will have one or more switching layers which can switch between communication configurations.*

An efficient design of the switching layer will allow to update its routing configuration partially or whole after a triggering event. Depending on the configuration, some functions will be employed while others are inactive.

If bi-directional communication channels are involved then we need to understand functions as bi-directional dynamical mechanisms as understood by Active Perception theory [82].

Active perception has many interesting aspects to it such as high contextuality of input during system activities and energetic efficiency. However, most electronic devices lack equipment where this approach could be employed down to physical interfacing. In the short run, we must therefore concentrate on organizing self-integrating environments around the classic concept of functions which are having distinct input and output interfaces.

## 12.2  The User-Desired Function F

In order to understand better how self-configuration, self-integration and error correction can be understood together, it is helpful to think of the IoT environment as a system for providing  an extensible set of user-desired functions $F_i$ to the user where $i$ is a natural integer. In figure 51 a user-desired function $F$ is shown and one level of its decomposition. This view is purely exemplary for us here. The key insight is that there are flows (red, orange) inside $F$ which are not visible and normally of no interest to the user of $F_i$.

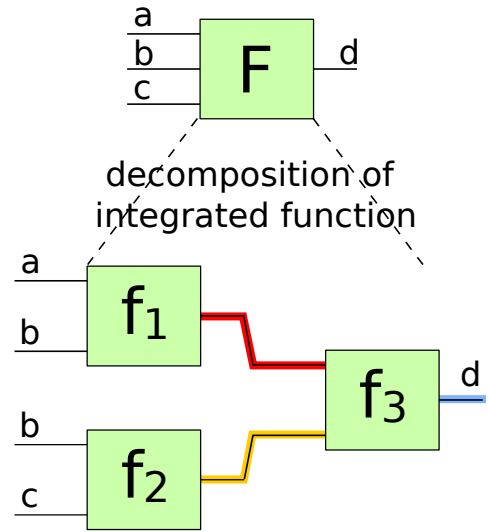

*Figure 51: A particular (dynamic) function is comprised of connected partial functions. F is a super system and $f_x$ are subsystems: Model-based systems engineers will find this idea straight forward because they often use signal-flow models like the one above.*

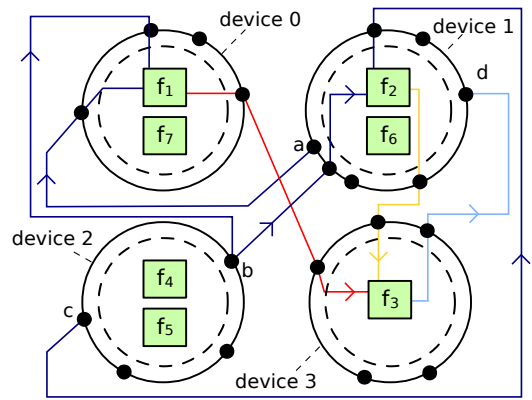

*Figure 52:Mapping of functions or subsystems to devices and mapping of flows to communication interfaces.*

The key chore in self-integration is to assemble $F$ like is shown in figure 52 (example). This can be done in a mix of two approaches: The first one is by self-corrective object interaction and the other is by user intervention whereby the goal is to reduce user intervention as much as possible. However, since technical devices of today lack a tremendous amount of knowledge about their intended uses, their design and because they are equipped with almost no features to rediscover this, integration of technical environments into new super-systems has been almost the sole responsibility of human users or engineers. Hence, in this work I concentrate on system-level concepts for how devices could rediscover their role in super-systems.

## 12.3   Integration in Managed Environments

There are two main approaches how this could be realized. The first approach is to explore new integrated functions $F_i$ based on combinations of available components and compatibility of interfaces between them. The user is presented with the function $F_i$ and decides if he wants to keep it. Since the user does not want to face new functions all the time, he must have means to tell the environment that he desires support in the form of a new "idea" or "solution proposition". This results in a user-triggered dialog between a "co-creative" environment and a "creative" user.

For example, the environment could understand some very basic concept like "activation" or "regulation" and find a communication path between user-indicated control and controlled device. An environment capable of updating the control device's interface and to manage the routing of signals to the controlled device would be of really much help in many scenarios.

Such basic interactive capabilities and predefined elementary concepts seem necessary because proposition of new integrated functions based on compatibility of interfaces alone is not likely to produce useful functions. Almost all practical applications demand from engineers that at some point they adapt incompatible interfaces in order to per-form a useful integrated function. In order to understand which communication path is critical, and hence eligible to adaptation, will require the a priori knowledge of the integrated function $F_i$.

This brings us to the second approach where $F_i$ is given and the environment shall restore it. Such restorations represent system reconfigurations [10] and will imply adaptation strategies between components.

Such *adaptive restoration* is necessary because either the user did perform a contradictory[16] function $F_j$ before $F_i$ or because the composition of the environment has changed – for example devices have been removed, replaced with similar ones or new devices were introduced providing new building blocks.

In figure 53 the problem is shown: Two devices were found which are offering only functions $f^*_1$ and $f^*_3$ which are a near match for $f_1$ and $f_3$. Involvement of these functions is the result of a discovery process in which the exact matches for $f_1$ and $f_3$ could not be found. Selection of nearly matching components will introduce a communication mismatch that remains unresolved unless there is a powerful technical approach able to "heal the link". Knowing which links should be "healed" is of great practical importance as users would normally not want that their devices show unexpected interventions, interactions or otherwise communications.

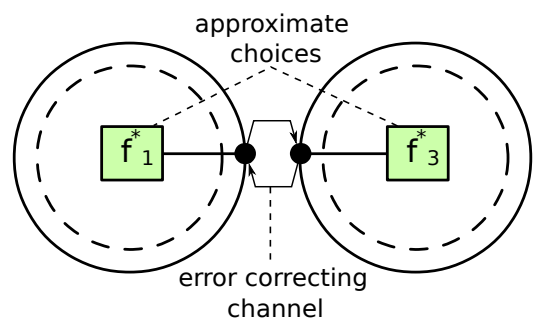

*Figure 53:candidates chosen with approximate functions of $f_1$ and $f_3$. Error correcting channels help to make the substitute configuration work.*

---

16  The function $F_j$ is using the same resources as $F_i$ but uses them to perform a different goal. Switching between $F_i$ and $F_j$ will result in reconfiguration of communications between resources and/or selection of other behavior modes.

In order to support selective "communications healing", the system representing or managing the "environment" and devices dwelling within it must be able to delimit and identify a particular function (or "functional context") in such a manner that it poses the same context to all relevant devices.

Additionally, since devices must not only enable a particular function (at a given time) but also prevent service to other concurring or interfering functions, it is necessary that devices can model in which communication context they are serving their functions. This would require that devices never perform a function outside a named context and that these contexts represent particular configurations of the switching layer at minimum.

### 12.4    Providing Knowledge to Error Correction Mechanisms

Once nearly matching functions are identified the task of error correction is to minimize any systematic error and to install error detection and correction functions for any remaining incompatibility.

This can be done in a general purpose and in a special purpose manner.

The general purpose correction will monitor communication links for their built-in error events and report it to the user but also the user should have access to a list of communications performed by the product and tell it if these communications were successful and which kinds of problems they experienced. At minimum this information will be collected in natural language and sent to the product developer in order to request a correction of the problem. General purpose correction would also involve general purpose techniques such as code recognition using pattern recognition techniques.

Special purpose techniques require additional design knowledge in order to identify and fix a particular problem. For example, a problem with quantity scaling could be fixed by rescaling – how much and which type of rescaling (linear or non-linear, filtered or not, etc.) will rely on two things: A diff-analysis between original design coupling and current design coupling and a knowledge base where certain types of diffs are associated with general solution strategies, eventually with pointers to software which could be downloaded and inserted into the communication path in order to fix the problem.

Supporting communications between devices and OEMs and between environmental configuration versions requires a data container which is living in the network and which can be used to store a precise history of design evolution and design performance. Based on this history it should be possible to reuse and alter communication channel interfaces in a deeply knowledgeable way, based rather on design purpose than on superficial designs' interface compatibility.

### 12.5    Using the Bubble from the Technical Interoperability Concept for the Advanced Industrial Interoperability Layer for Providing Design Knowledge to Error Correction Mechanisms

For exactly this process the *Bubble* has been proposed in [79] and in [83]. Its job is to hold and reuse design data in context of use from design time to runtime. It can be used as an object representation, as a system representation or as a communication context, but will require access to a Bubble-organized environment as shown in figure 54. Such Bubble-organized environment would be a natural repository for storing configurations of devices because Bubbles support basic operations which are also applicable to physical objects such as replication, instantiation and variation.

This environment architecture assumes that there is a local IoT administration authority which is responsible for providing the right device configurations for the right functions at the right time. This administrative unit is called "user-controlled environment manager" (UCEM) in figure 54. It monitors presence of user-devices in its scope of administration, usually sufficiently well-defined by connectivity to the Intranet's networks, and it offers an integrated interface for the user. The user will use device interfaces to override current environment configurations or the system will "borrow" devices' user interfaces in order to communicate conveniently with the user by using available devices in user's physical proximity. The UCEM can observe if the user is changing device configurations towards a known environment configuration and propose something like an "auto-complete" for the configuration of the environment.

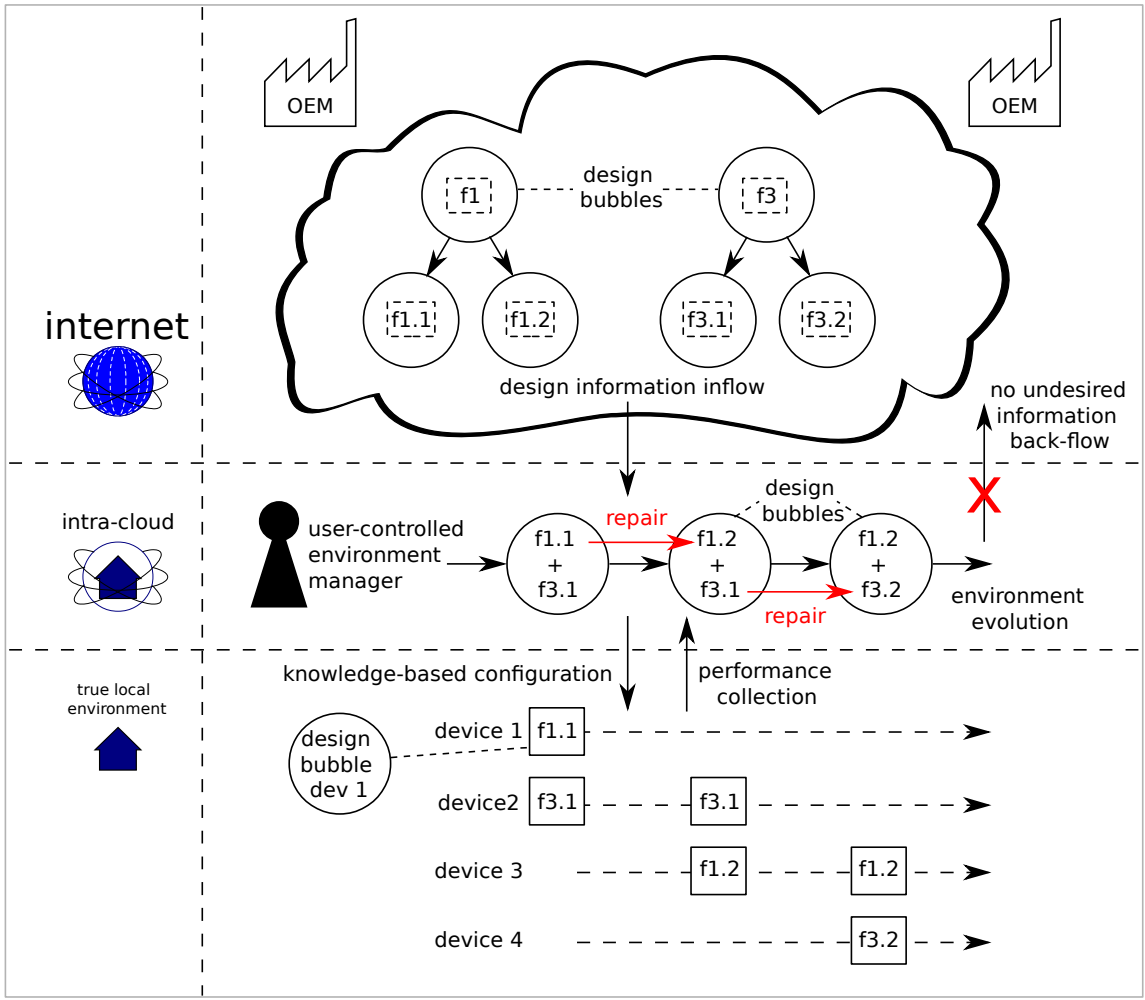

*Figure 54: In a bubble-enabled environment it is possible to understand coming and leaving resources not so much based on their interfaces ("superficial understanding") but on design evolution relationships and additional design description information ("deep understanding"). This should allow to deploy a function Fi in the true local environment in such a manner that the amount of communication repair is minimized. Error correcting features can be deployed for desired communication links only.*

Usually, an authority like the UCEM knows nothing about the devices except for the exposed interfaces. In a bubble-enabled infrastructure this is different. It can ask the device for its unique CAGUIDS[17] identifier which is globally unique, much like MAC address. This identifier can be used on the Internet in order to find OEM-published design descriptions, potentially suitable for machine reasoning and machine processing. These descriptions can be used by the UCEM in order to conclude that there exist devices with related functionality in its scope. It can at least understand the function of new objects by analogy in previous uses but could require some adaptation in configuration and communication. This means that the UCEM can rely on OEM's design description and on its own experience with the type of design in question in order to conclude purpose and setup of device. For this it can relate to a permeating design ontology system organized of machine processable knowledge boxes, the Bubbles.

17 Cascading Global Unique Identifier System

## 12.6 Objects with Unknown Communication Friends

In section 10 the issue of understanding functional and dysfunctional behaviors was considered crucial for deciding which behaviors could be interpreted as "functional" and which are "abnormal enough" in order to be useful for use in new communications. Making such assertions could be accelerated if additional knowledge was available in a common formalism suitable to computer processing and which is easily accessible in relation to the objects in range of interaction.

The task to identify potential communicative behaviors and modes of operation could be simply provided in the design containers, the Bubbles, which can be obtained via Internet/Intranet connections. In fact, when one object detects another object and is attempting to locate visible Bubble contents via its Wi-Fi or Ethernet connection, it could be directly routed to the object of interest and talk to its Bubble governor. Devices could be equipped with visual markers and network broadcasting protocols in order to announce themselves and which Bubble is responsible for their design. In ideal scenarios Bubbles are exposing the complete set of design details which can be used to address a particular function of the device. The more reuse is practiced among devices, the higher the general probability that devices will find functions and resources on remote objects which they are already familiar with and know how to use them – contemporary devices can only make implicit assumptions about this based on detected device interfaces.

Access to design data will involve reasoning activities about designs in new communication contexts. Clearly, many sophisticated consumer electronics have sufficient computing power and storage to perform reasoning on the device but they could also delegate complex reasoning activities to knowledge reasoners on the web. For this only references to Bubbles of interest must be transmitted. If parts of the design Bubble are already available on the Internet, then reasoners will not rely on thin network bandwidths to the devices alone.

## 12.7 Harnessing Global Design Data Access without Risk of Surveillance or Big Data Collecting Infrastructure

The total system configuration needs not to be stored or restored as a unit as the configuration stretches several architectural layers of the overall system. At the lowest level a configuration can describe a network port configuration and at the highest level a configuration can describe devices included in it, which modes they should be in and which communications shall take place. It makes sense to keep device-specific configuration information on the device and not to force the supra-management layers to have to collect and manage all kinds of types of detailed configuration data. In fact, devices can (and should) maintain own configurations used in super-configurations rather than exposing all internal parameters to external managers (as is usually proposed for many contemporary IoT architectures in order to provide OEMs with the opportunity to collect large amounts of data).

This is a privacy feature but it is also allowing devices to come up with a "replacement" configuration for the same purpose if, for example, the device experiences failures - but it can only do so if the function it performs is represented by an object with explicitly defined interfaces and function constraints – it "knows what it does" so to say. This description can and must come from external management systems but they have to have published objects to which they can "attach" their expectations in form of snippets of standardized description language. For example, if a device realizes that its main input is defective and needs to redirect this input source to available input alternatives then it must be able to check if this new input can satisfy signal resolution requirements. If the required input is Boolean then switching from one flip-button to another is implementable. Of course, more complex reallocation scenarios are possible to imagine and I see no limits to how complex these scenarios can become.

The above mentioned fascinating capabilities require computational concepts which are clearly exceeding most contemporary device designs. Most devices, if involved in a re-configurable system, are configured remotely. A remotely configured object will rely on high-bandwidth links to a centralized control authority and sufficient intelligence designed into it. This is the current state of affairs.

However, if devices provide such contextual device configuration objects then we can get not only locally administered device resilience but also lower demand for channel capacity, improved privacy and easier life-cycle management of device configuration data (e.g. data is destroyed along with the device).

Importance of reduced bandwidth requirements should not be underestimated: Reducing bandwidth requirements is critical if devices shall exploit non-IP channels for communication which are a lot slower than Wi-Fi or Ethernet networks but also a lot more trusted due to physical localization of communications.

## 12.8    Avoiding Errors in Communications by Reusing Context-aware Device Configurations

If devices can manage their own configurations then they can also optimize shared function among different environments. For example, if a device is moved between environments A and B where it contributes to integrated functions $F_A$ and $F_B$ in the same fashion then it can consolidate the configuration into a single one without administrative intervention of environment A or B. More realistically, once this device moves into environment B after being in environment A, it will want to reuse the configurations developed in environment A. This original clone should be as thin as possible, i.e. require as little additional memory as possible. If environments A and B start to diverge and start to incrementally diverge the required configuration on the device, the device will have to manage the configuration divergence in a smart manner. In terms of error correction such reuse practice implies that communication solutions are also reused without a complete re-reasoning and re-evaluation of joint design and need only be adapted whenever needed.

## 12.9    The Adapter Middle-Layer

Hitherto we have seen a basic communication channel concept as shown in figure 34 which shall be an architectural framework for designing powerful error correcting channels for low level and high level types of content as shown in figure 9. In all concepts, error correction requires redundancy in order to allow reasoning about what is the correct interpretation of data. This redundancy can be provided with the stream, can be accumulated, can be provided externally or can be actively requested – leading us to a merged concept of regulation and communication. A communication channel with feedback is described as two-edged channel (abbreviated as 2e) and a bi-direction 2e channel is described as a four-edged channel (abbreviated as 4a). In a 4e channel no information is conveyed as long as emitted signal modulations are consistent with a given error model. The information modulation in a channel leads to a cease of of observable signal.

In figure 44 a semi-technical interpretation of four edged, nested communications was proposed but the question remained where in a device architecture this should be understood. In figure 50 a paradigm was proposed in which the idea is that a device has a senso-motoric "hull" which is interacting with intern functions via a routing and prioritization layer.

In general, this work proposes to use two sources of error correction knowledge within this context.

The first is to use soft computing techniques, particularly from pattern recognition and machine learning, in order to map and trigger behaviors known to be "dysfunctional" to basic internal concepts relevant to device communication.

The second is to use artificial intelligence techniques, such as design reasoning engines and communication planners, both on and off device, to provide adaptive connectivity between functions.

The question is then how these features could be combined in a software architecture?

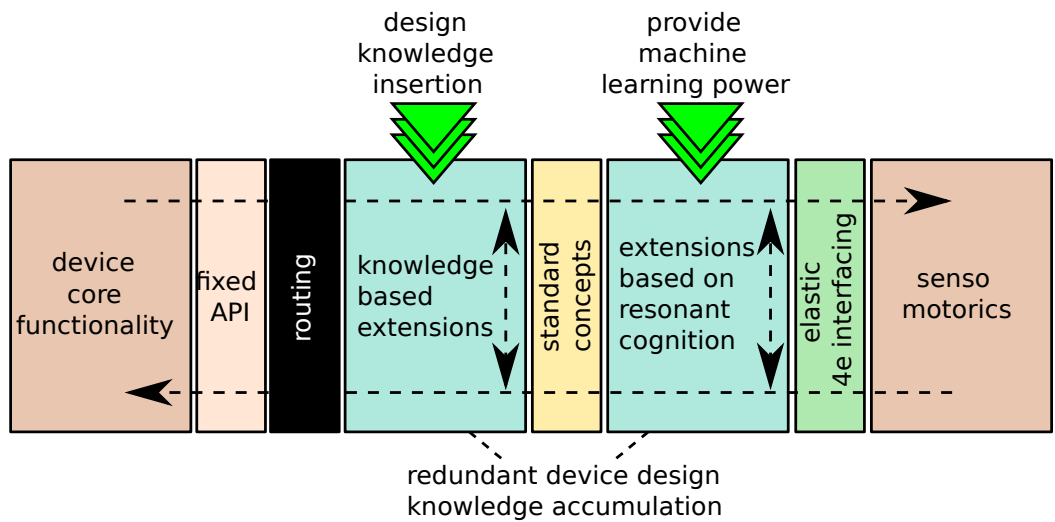

*Figure 55: The communication driver prepared for extended error correction will provide extension points for its mapping and conversion mechanisms for two types of correction knowledge sources. The first source is external knowledge provided for example via Internet and a design Bubble and the second source is a machine learning environment suitable for detecting and optimizing codes. The two sources interact via a greatly simplified interface layer with universal standard concepts.*

Figure 55 is making a proposition for this: On the left we find the fixed core of functionality found in the device which is configured via a routing layer to the hull (right side). The number of such routings can be any and communications must provide the addressing information along with the payload. Since addresses travel along with the payload and may be treated as such, recognizing and defining addresses for other parties can be performed with same set of methods, provided that information can be accumulated in middle layers and information is pushed forward only if sufficient addressing precision was reached. Ultimately, all payload can be understood as addresses to nested functionality.

Data produced and consumed at this stage is quite technical and requires a sophisticated degree of technical precision which I don't believe can be sufficiently provided with soft-computing techniques. Creating corrective shell around the fixed part of functionality will be easier if a relatively technical and explicitly formulated design description was provided to a reasoning layer which can rely on knowledge it has accumulated from communications and external sources (such as the Bubblenet). This layer is excellent for providing challenging technical concepts such as time-outs, synchronization, transactions, alge-braic transformations, etc.

In order to transform the concepts used by this layer into sequences of active perception I propose to insert a layer of standardized basic concepts such as activations, controls, directions, conditions, constraints, etc. Those can be mapped to external interactive behaviors (active perception) by on-device learning. Frankly, the learning needs not to occur on the device but if the learning has to be externalized then the device must collect and forward large quantities of senso-motoric data to an external service which is capable to train new models for the device – this could be a privacy issue.

These learned active perception models will involve information modulation like shown in figure 44 with the partially filled "boxes". The job of the 4e channel driver is to define when and how to transmit data given a particular interaction device (sensor, motor, mixed).

# 13 Discussion and Conclusions

## 13.1 Matter of Interest

This work is interested in investigating conceptual design aspects for devices which shall contribute to a self-integrating environment and environments fit for the fabrication web accessible over the Internet.

Characteristic for such environments is that they are not well defined. Many independent parties contribute to hardware and software of the environment. IoT and I4 devices must handle a great variety of environmental configurations and updates of any of its parts without prior notice.

Such updates can corrupt function by rendering communications incompatible. Communication symbols could have changed semantics or the channels have lowered reliability. It is expected from devices that they provide means for the user to fix at least basic problems or that devices can correct communications themselves.

## 13.2 Going Beyond Classic Error Correction Concepts

This paper is embracing systematic and random errors in one concept as systematic errors can appear as random error sources. There is no dichotomy between systematic error and random error but rather a continuum. Modern device resilience concepts should propose architectures which are capable to adequately handle problems anywhere in this spectrum. In the end, randomness is defined as the inability to explain or predict observations. By this definition, systems experiencing random errors can in truth suffer from systematic errors for which they only have no model. An error correction model should therefore include propositions how to acquire such models in order to reduce the remaining random element as much as possible.

When communications on traditional paths fail the question arises if devices (or more generally *intelligent objects*) can compensate failing channels with new ones. For this purpose device should be able to detect and trigger behaviors which are not purely motivated by function. I admit that recognizing functional behaviors from communicative "dysfunctional" behaviors is a complex cognitive task as the observer model must attempt to understand what is the purpose of other objects and what are their regular strategies to satisfy it.

## 13.3 Error Correction using Design Knowledge from the Bubblenet

Because of this challenge I have bridged over to Design Bubbles from the Technical Interoperability Concept for the Advanced Industrial Interoperability Layer. Bubbles were designed to manage design data in engineering environments, as sophisticated method for design reuse, configuration and variant management across a wide range of sites. Ubiquitous access to the *Bubblenet* can greatly reduce the time and energy required to model and understand other objects in physical proximity, notably even if they have no electronics. For example, a robot vacuum cleaner could detect a chair, detect an ID to the Bubblenet on it, and then acquire some physical descriptions for this chair. It could use this information in order to decide, if it can push it or not or if the object's owner has defined special treatment for this particular object, such as a safety radius around it. Such information can be considered as "error prevention" between an intelligent object and a passive object. It be reminded that passive objects also have a "communicative hull" where all channels are highly resistive, i.e. they will not actively respond. This is a special case in the more general case where both parties can be intelligent objects and try to influence each other.

It has been emphasized that objects/devices are not necessarily in permanent communication constellations and that objects/devices are moved between constellations sharing a common design history. Hence by supporting on-device communications configuration management and reuse, devices can gain the ability to quickly switch between modes of operation/communication, thus reducing incompatibility time and thus error time. The use of on-device Bubbles is proposed for this job. In sum all the configuration and performance data in the on-board Bubbles will constitute a useful self-model of the device.

### 13.4 Supporting Local Device Plants in Intra-Clouds

In the more interesting unconstrained communication cases we can assume adaptive modeling capabilities in all involved objects. They are using them to find out about other objects in their environment and which behaviors are characteristic for their task, which of them can be exploited and which ones should be suppressed. Because systems can learn to communicate with each other in terms of suppressing each other, it is important to think about how devices in an environment can act in a well balanced way.

Since devices with adaptive error correction fall under the category of resilient self-configuring machines, their capabilities contribute a difficult to control element to the overall functional system configuration. From this results a need to define basic behaviors which renders them more generously collaborative on average. Such could be the ability to postpone actions for defined time windows if the environment is appearing to actively process something. If devices are only designed to seek own best functional solution under all known environmental conditions then we could observe race conditions between devices, i.e. devices trying to gain domination over other devices (cf. fig. 5). During race conditions devices will create policies in order to neutralize other object's strategies and to enforce success of own missions with the effect they produce more and not less cost in the overall system. This behavior is the result of the use of knowledge about objects for deriving active interaction policies with these objects under the constraint of their stated mission. This could lead to overall poor electronics environment performance.

Much of such resource conflict can be simplified in resolution if devices obey a central UCEM (user-controlled environment manager) which is sitting on site for privacy reasons. If UCEM deactivates or reconfigures objects in its scope then this should prevent modeling of harmful interaction protocols which would also be made "robust" just like any synergetic communication is made robust against external interference using adaptive error correction strategies.

### 13.5 Main Concepts

This paper is proposing to think of error correction not mainly as a mathematical discipline for defining error correction codes but to recognize that:

- all non-binary contents can rely on redundant information in order to be repaired.

- error correction strength can be balanced among nested protocols.

- redundancy in the signal is not the only source of error correction information. Such information can be collected by modeling or provided as knowledge units on different paths.

- error correction can imply activity from the receiver.

- error correction and control theory are intimately related.

- spatio-temporal codes and error resisting modulation can be co-evolved between objects

- success of IoT devices will rely on intelligent merging of Ethernet-based and locally physical interactions.

- adaptive error correction demands from device an architectures where internal device communication paths can be re-wired

### 13.6 Future Work

Despite the length of this paper and reference to literature, the concept of integrated error presented here lack sufficient experimental validation or theoretical formalisms. It will be therefore left to future work to demonstrate implementations and theoretical frameworks for integrated adaptive error correction.

# References

[1]     J. Gubbi, R. Buyya, S. Marusic, and M. Palaniswami, "Internet of Things (IoT): A vision, architectural elements, and future directions," *Futur. Gener. Comput. Syst.*, vol. 29, no. 7, pp. 1645–1660, 2013.

[2]     J. Lee, B. Bagheri, and H. Kao, "A Cyber-Physical Systems architecture for Industry 4.0-based manufacturing systems," *Manuf. Lett.*, vol. 3, pp. 18–23, 2015.

[3]     I. Chatzigiannakis, H. Hasemann, M. Karnstedt, O. Kleine, A. Kröller, M. Leggieri, D. Pfisterer, K. Römer, and C. Truong, "True self-configuration for the IoT," *Proc. 2012 Int. Conf. Internet Things, IOT 2012*, pp. 9–15, 2012.

[4]     T. K. Moon, *Error Correction Coding*. New Jersey: Wiley-Interscience, 2005.

[5]     L. Kung, "Introduction to Error Correcting Codes," 2016.

[6]     J. M. Cioffi, "Signal Processing and Detection," *Lect. notes course EE379A Digit. Commun. - Signal Process.*, pp. 1–101, 2014.

[7]     M. C. Huebscher and J. A. Mccann, "A survey of Autonomic Computing — degrees, models and applications," vol. V, pp. 1–31, 2002.

[8]     M. Parashar and S. Hariri, "Autonomic Computing: An Overview," pp. 247–259, 2005.

[9]     O. Brukman, S. Dolev, Y. Haviv, and R. Yagel, "Self-stabilization as a foundation for autonomic computing," *Proc. - Second Int. Conf. Availability, Reliab. Secur. ARES 2007*, pp. 991–998, 2007.

[10]    A. Lodwich, "Exploring high-level Perspectives on Self-Configuration Capabilities of Systems," p. 45, 2016.

[11]    A. Note, "Agilent Digital Modulation in Communications Systems — An Introduction Application Note 1298 Introduction," pp. 1–5, 2005.

[12]    A. R. Hammons and H. E. Gamal, "On the theory of space-time codes for PSK modulation," *IEEE Trans. Inf. Theory*, vol. 46, no. 2, pp. 524–542, 2000.

[13]    V. Chen, E. Grigorescu, and R. D. E. Wolf, "Error-correcting data structures ∗," *SIAM J. C OMPUT .*, vol. 42, no. 1, pp. 84–111, 2013.

[14]    B. Sklar, "Fundamentals of Turbo Codes by," *Prentice Hall Mar*, pp. 1–30, 1993.

[15]    C. Castelfranchi, "Silent agents: From observation to tacit communication," *Proc. Adv. Artif. Intell.*, vol. 4140, pp. 98–107, 2006.

[16]    P. R. Cohen and H. J. Levesque, "Rational interaction as the basis for communication," techreport, 1988.

[17]    M. L. L. David H. Ackley, "Altruism in the Evolution of Communication," in *Artificial Life IV*, Rodney Brooks and Pattie Maes, Ed. MIT CogNet, 1994.

[18]    J. Noble, "Cooperation, conflict and the evolution of communication," *Adapt. Behav.*,

vol. 7, no. 3–4, p. 349, 1999.

[19]  F. Heylighen, "Self-organization in communicating groups: The emergence of coordination, shared references and collective intelligence," *Underst. Complex Syst.*, pp. 117–149, 2013.

[20]  R. Lundh, "Robots that Help Each Other: Self-Configuration of Distributed Robot Systems," p. 205, 2009.

[21]  R. Lundh, L. Karlsson, and A. Saffiotti, "Dynamic self-configuration of an ecology of robots," in *IEEE International Conference on Intelligent Robots and Systems*, 2007, pp. 3403–3409.

[22]  W. Sheng, Q. Yang, J. Tan, and N. Xi, "Distributed multi-robot coordination in area exploration," *Rob. Auton. Syst.*, vol. 54, no. 12, pp. 945–955, 2006.

[23]  Z. Yan, N. Jouandeau, and A. A. Cherif, "A survey and analysis of multi-robot coordination," *Int. J. Adv. Robot. Syst.*, vol. 10, 2013.

[24]  S. Nolfi, "Emergence of communication in embodied agents: co-adapting communicative and non-communicative behaviours," *Conn. Sci.*, vol. 17, no. 3–4, pp. 231–248, 2005.

[25]  E. A. Di Paolo, "An investigation into the evolution of communication," *Adapt. Behav.*, vol. 6, no. 2, pp. 285–324, 1997.

[26]  L. Steels, "Evolving Grounded Communication for Robots," *Trends Cogn. Sci.*, vol. 7, pp. 308–312, 2003.

[27]  T. C. Scott-Phillips, R. A. Blythe, A. Gardner, and S. A. West, "How do communication systems emerge?," *Proc. R. Soc. B Biol. Sci.*, vol. 279, no. 1735, pp. 1943–1949, 2012.

[28]  J. Noble and L, "The Evolution of Animal Communication Systems: Questions of Function Examined through Simulation," *Univ. Southampt. Res. Repos. ePrints Sot.*, p. I–XX, 1-195, 1998.

[29]  F. Heylighen, "Stigmergy as a universal coordination mechanism II: Varieties and evolution," *Cogn. Syst. Res.*, vol. 38, pp. 50–59, 2016.

[30]  T. Page, "Usability of text input interfaces in smartphones," *J. Des. Res.*, vol. 11, no. 1, pp. 39–56, 2013.

[31]  X. A. Chen, T. Grossman, and G. Fitzmaurice, "Swipeboard: A Text Entry Technique for Ultra-Small Interfaces That Supports Novice to Expert Transitions," *Proc. UIST '14*, pp. 615–620, 2014.

[32]  A. Quattrini Li, L. Sbattella, and R. Tedesco, "PoliSpell: An Adaptive Spellchecker and Predictor for People with Dyslexia," in *User Modeling, Adaptation, and Personalization: 21th International Conference, UMAP 2013, Rome, Italy, June 10-14, 2013 Proceedings*, S. Carberry, S. Weibelzahl, A. Micarelli, and G. Semeraro, Eds. Berlin, Heidelberg: Springer Berlin Heidelberg, 2013, pp. 302–309.

[33]    M. Jones, "Contextual spelling correction using latent semantic analysis," *Proc. fifth Conf.*, pp. 166–173, 1997.

[34]    Y. Bassil and P. Semaan, "ASR Context-Sensitive Error Correction Based on Microsoft N-Gram Dataset," *arXiv Prepr. arXiv1203.5262*, vol. 4, no. 1, pp. 34–42, 2012.

[35]    Y. Mohapatra, A. K. Mishra, and A. K. Mishra, "Spell Checker for OCR," vol. 4, no. 1, pp. 91–97, 2013.

[36]    M. Jeong, B. Kim, and G. G. Lee, "Using Higher-level Linguistic Knowledge for Speech Recognition Error Correction in a Spoken Q/A Dialog," *Proc. HLT-NAACL 2004 Work. Spok. Lang. Underst. Conversational Syst. High. Lev. Linguist. Inf. Speech Process.*, pp. 48–55, 2004.

[37]    A. Rozovskaya and D. Roth, "Generating confusion sets for context-sensitive error correction," *Proc. 2010 Conf. Empir. methods Nat. Lang. Process.*, no. October, pp. 961–970, 2010.

[38]    N. Nagarajan and M. Pop, "Sequence assembly demystified," *Nat Rev Genet*, vol. 14, no. 3, pp. 157–167, Mar. 2013.

[39]    D. Martin, D. Martin, M. Paolucci, M. Paolucci, S. McIlraith, S. McIlraith, M. Burstein, M. Burstein, D. McDermott, D. McDermott, D. McGuinness, D. McGuinness, B. Parsia, B. Parsia, T. Payne, T. Payne, M. Sabou, M. Sabou, M. Solanki, M. Solanki, Others, and Others, "Bringing semantics to web services: The OWL-S approach," *Proc. First Int. Work. Semant. Web Serv. Web Process Compos. SWSWPC 2004*, vol. 3387, pp. 26–42, 2005.

[40]    C. S. Quan Z. Sheng, Xiaoqiang Qiao, Athanasios V. Vasilakos and X. X. Scott Bourne, "Web services composition: A decade's overview," *Inf. Sci. Informatics Comput. Sci. Intell. Syst. Appl.*, vol. 280, pp. 218–238, 2014.

[41]    A. Lodwich, "Note on Communication Incompatibility Types," 2016.

[42]    D. A. Ferrucci, D. C. Gondek, and W. W. Zadrozny, "Error correction using fact repositories." Google Patents, 2013.

[43]    R. Narasimha and N. Shanbhag, "Design of energy-efficient high-speed links via forward error correction," *IEEE Trans. Circuits Syst. II Express Briefs*, vol. 57, no. 5, pp. 359–363, 2010.

[44]    D. Y. and Q. L. Zhen Tian and Z. Tian, "Energy efficiency analysis of error control schemes in wireless sensor networks," *Wirel. Commun. …*, vol. 9, no. 60672036, pp. 401–405, 2008.

[45]    A. Elyas, M. N. B. Zakaria, H. Yosif, and S. B. Ibrahim, "An Efficient Energy Adaptive Hybrid Error Correction Technique for Underwater Wireless Sensor Networks," vol. 5, no. 3, pp. 1389–1395, 2011.

[46]    N. A. Alrajeh, U. Marwat, B. Shams, S. Saddam, and H. Shah, "Error Correcting Codes in Wireless Sensor Networks : An Energy Perspective," vol. 818, no. 2, pp. 809–818,

2015.

[47] D. Bull, S. Das, K. Shivashankar, G. S. Dasika, K. Flautner, and D. Blaauw, "A power-efficient 32 bit ARM processor using timing-error detection and correction for transient-error tolerance and adaptation to PVT variation," *IEEE J. Solid-State Circuits*, vol. 46, no. 1, pp. 18–31, 2011.

[48] F. Dehmelt, "Adaptive ( Dynamic ) Voltage ( Frequency ) Scaling — Motivation and Implementation," no. March, pp. 1–10, 2014.

[49] M. S. Diggs and D. E. Merry, "Storage subsystem capable of adjusting ECC settings based on monitored conditions." Google Patents, 2012.

[50] S. A. Hussain, T. Shah, N. A. Azam, A. A. De Andrade, and A. N. Malik, "Spectrum Distribution in Cognitive Radio: Error Correcting Codes Perspective," *Int. J. Distrib. Sens. Networks*, vol. 2014, p. Article ID 864916, 2014.

[51] T. W. Rondeau, A. B. Mackenzie, J. H. Reed, S. F. Midkiff, and S. B. Ball, "Application of Artificial Intelligence to Wireless Communications," *Artif. Intell.*, vol. Doctor of, 2007.

[52] P. W. Dent, "Adaptively self-correcting modulation system and method." Google Patents, 1994.

[53] L. J. O'Donnell, H. M. Hagberg, and M. A. Gorman, "Adaptive error correction." Google Patents, 2010.

[54] D. Martinez, T. Hengeveld, and M. Axford, "Adaptive error correction for a communications link." Google Patents, 1999.

[55] E. Ayanoglu, R. D. Gitlin, T. F. L. Porta, S. Paul, and K. K. Sabnani, "Adaptive forward error correction system." Google Patents, 1997.

[56] P. . J. M. Havinga, "Energy efficiency of error correction on wireless systems," *WCNC. 1999 IEEE Wirel. Commun. Netw. Conf. (Cat. No.99TH8466)*, 1999.

[57] A. V. Aho and T. G. Peterson, "A minimum distance error-correcting parser for context-free languages," *SIAM J. Comput.*, vol. 1, no. 4, pp. 305–312, 1972.

[58] K. S. Fu, "Error-Correcting Parsers for Formal Languages," *IEEE Trans. Comput.*, vol. C-27, no. 7, pp. 605–616, 1978.

[59] M. De Jonge, E. Nilsson-Nyman, L. C. L. Kats, and E. Visser, "Natural and flexible error recovery for generated parsers," in *Lecture Notes in Computer Science (including subseries Lecture Notes in Artificial Intelligence and Lecture Notes in Bioinformatics)*, 2010, vol. 5969 LNCS, pp. 204–223.

[60] R. Grush, "The emulation theory of representation: Motor control, imagery, and perception," pp. 377–442, 2004.

[61] A. E. Mohr, E. A. Riskin, and R. E. Ladner, "Unequal loss protection: graceful degradation of image quality over packet erasure channels through forward error

correction," *IEEE J. Sel. Areas Commun.*, vol. 18, no. 6, pp. 819–828, 2000.

[62]   C. B. and L. C. P. Schlegel, *Trellis and turbo coding*. Piscataway, NJ, USA: Wiley-Interscience, 2004.

[63]   M. Wick, M. Ross, and E. Learned-Miller, "Context-Sensitive Error Correction: Using Topic Models to Improve OCR," *Ninth Int. Conf. Doc. Anal. Recognit. (ICDAR 2007) Vol 2*, pp. 1168–1172, 2007.

[64]   Y. Labrador, M. Karimi, and D. Pan, "Modulation and Error Correction in the Underwater Acoustic Communication Channel," *IJCSNS Int. J. Comput. Sci. Netw. Secur.*, vol. 9, no. 7, pp. 123–130, 2009.

[65]   M. Franceschetti and P. Minero, "Elements of Information Theory for Networked Control Systems," in *Information and Control in Networks*, G. Como, B. Bernhardsson, and A. Rantzer, Eds. Cham: Springer International Publishing, 2014, pp. 3–37.

[66]   P. Minero, M. Franceschetti, S. Dey, and G. N. Nair, "Data rate theorem for stabilization over time-varying feedback channels," *IEEE Trans. Automat. Contr.*, vol. 54, no. 2, pp. 243–255, 2009.

[67]   W. S. Wong and R. W. Brockett, "Systems with finite communication bandwidth contraints -- {II}: {S}tabilization with limited information feedback," *IEEE Trans. Automat. Contr.*, vol. 44, no. 5, pp. 1049–1053, 1999.

[68]   M. T. Motley, "On whether one can (not) not communicate: An examination via traditional communication postulates," *West. J. Commun. (includes Commun. Reports)*, vol. 54, no. 1, pp. 1–20, 1990.

[69]   J. B. Bavelas, "Behaving and communicating: A reply to Motley," *West. J. Speech Commun.*, vol. 54, no. 4, pp. 593–602, 1990.

[70]   S. Álvarez-García and N. Brisaboa, "Compressed k2-triples for full-in-memory RDF engines," *arXiv Prepr. arXiv …*, pp. 1–9, 2011.

[71]   M. Z. Jeff Z. Pan, José Manuel Gómez Pérez, Yuan Ren, Honghan Wu, Haofen Wang, "Graph Pattern Based RDF Data Compression," *Lect. Notes Comput. Sci. (including Subser. Lect. Notes Artif. Intell. Lect. Notes Bioinformatics)*, vol. 8943, no. November, pp. 239–256, 2015.

[72]   J. Fernández, "RDF compression: basic approaches," *Proc. 19th …*, pp. 3–4, 2010.

[73]   K. Kukich, "Technique for automatically correcting words in text," *ACM Comput. Surv.*, vol. 24, no. 4, pp. 377–439, 1992.

[74]   M. V. M. Mahoney, "Fast text compression with neural networks," *Proc. AAAI FLAIRS*, pp. 0–4, 2000.

[75]   J. Bruck and M. Blaum, "Neural Networks, Error-Correcting Codes, and Polynomials over the Binary n-Cube," *IEEE Trans. Inf. Theory*, vol. 35, no. 5, pp. 976–987, 1989.

[76]   A. Lodwich, "Differences between Industrial Models of Autonomy and Systemic Models

of Autonomy," Schwerte, 2016.

[77]  A. Lodwich, "How to avoid ethically relevant Machine Consciousness," Schwerte, 2016.

[78]  D. MacKenzie, "Compressed Sensing Makes Every Pixel Count," *Signal Processing*, vol. 7, no. July, pp. 114–127, 2007.

[79]  A. Lodwich and J. M. Alvarez-rodriguez, "Beyond Interoperability in Critical Systems Engineering," pp. 1–7.

[80]  W. Kang, Y. Zhang, Z. Wang, J.-O. Klein, C. Chappert, D. Ravelosona, G. Wang, Y. Zhang, and W. Zhao, "Spintronics: Emerging Ultra-Low-Power Circuits and Systems Beyond MOS Technology," *J. Emerg. Technol. Comput. Syst.*, vol. 12, no. 2, p. 16:1--16:42, Sep. 2015.

[81]  T. Tang, L. Xia, B. Li, R. Luo, Y. Chen, Y. Wang, and H. Yang, "Spiking Neural Network with RRAM: Can We Use It for Real-world Application?," in *Proceedings of the 2015 Design, Automation & Test in Europe Conference & Exhibition*, 2015, pp. 860–865.

[82]  R. Bajcsy, "Active Perception," in *Proceedings of the IEEE, Vol. 76, No. 8*, 1988, pp. 996–1005.

[83]  José Maria Alvarez-Rodríguez, Lodwich, Aleksander, "Bubbles: a data management approach to create an advanced industrial interoperability layer for critical systems development applying reuse techniques," 2016.

### 13.7   Links

**Exploring high-level Perspectives on Self-Configuration Capabilities of Systems**

https://lodwich.net/Science/self-configuring_systems.pdf

**How to avoid ethically relevant Machine Consciousness**

https://lodwich.net/Science/machine_consciousness.pdf

**Differences between Industrial Models of Autonomy and Systemic Models of Autonomy**

https://lodwich.net/Science/LevelsOfAutonomy.pdf